\documentclass[12pt]{article}

\usepackage{amssymb,amsmath,amsfonts,eurosym,geometry,ulem,graphicx,caption,color,setspace,sectsty,comment,footmisc,caption,natbib,pdflscape,subfigure,array,bm}
\usepackage{algorithmic}
\usepackage{booktabs, threeparttable, dcolumn, tikz, setspace, multirow, listings, rotating}
\usepackage[capposition=top]{floatrow} 
\usetikzlibrary{arrows,calc}
\usepackage{array}
\usepackage{xcolor}
\usepackage{chngcntr}
\usepackage[obeyspaces]{url}
\pdfpageattr {/Group << /S /Transparency /I true /CS /DeviceRGB>>} 

\DeclareMathOperator{\E}{\mathbb{E}}

\newcolumntype{L}[1]{>{\raggedright\let\newline\\\arraybackslash\hspace{0pt}}m{#1}}
\newcolumntype{C}[1]{>{\centering\let\newline\\\arraybackslash\hspace{0pt}}m{#1}}
\newcolumntype{R}[1]{>{\raggedleft\let\newline\\\arraybackslash\hspace{0pt}}m{#1}}

\newcommand\deltaDemand{20\% } 
\newcommand\deltaPUN{45\% } 
\newcommand\deltaRedispatchCost{73\% } 
\newcommand\deltaRedispatchCostDemand{108\% } 

\begin{document}

\begin{titlepage}
\title{\vspace{-2.0cm}(Machine) Learning from the COVID-19 Lockdown about Electricity Market Performance with a Large Share of Renewables}
\author{Christoph Graf\thanks{Department of Economics, Stanford University, Stanford, CA 94305-6072, cgraf@stanford.edu. Financial support from the Austrian Science Fund (FWF), J-3917, and the Anniversary Fund of the Oesterreichische Nationalbank (OeNB), 18306, is gratefully acknowledged.}
\and Federico Quaglia\thanks{Terna S.p.A., Viale Egidio Galbani, 70, 00156 Rome, Italy, federico.quaglia@terna.it.} \and Frank A.\ Wolak\thanks{Program on Energy and Sustainable Development (PESD) and Department of Economics, Stanford University, Stanford, CA 94305-6072, wolak@zia.stanford.edu.}}
\date{October 21, 2020}

\maketitle
\vspace{-1.2cm}
\begin{abstract}
\noindent The negative demand shock due to the COVID-19 lockdown has reduced net demand for electricity---system demand less amount of energy produced by intermittent renewables, hydroelectric units, and net imports---that must be served by controllable generation units. Under normal demand conditions, introducing additional renewable generation capacity reduces net demand. Consequently, the lockdown can provide insights about electricity market performance with a large share of renewables. We find that although the lockdown reduced average day-ahead prices in Italy by 45\%, re-dispatch costs increased by 73\%, both relative to the average of the same magnitude for the same period in previous years. We estimate a deep-learning model using data from 2017--2019 and find that predicted re-dispatch costs during the lockdown period are only 26\% higher than the same period in previous years. We argue that the difference between actual and predicted lockdown period re-dispatch costs is the result of increased opportunities for suppliers with controllable units to exercise market power in the re-dispatch market in these persistently low net demand conditions. Our results imply that without grid investments and other technologies to manage low net demand conditions, an increased share of intermittent renewables is likely to increase costs of maintaining a reliable grid.
\vspace{0.15in}\\
\noindent\textbf{Keywords:} Net demand shock; Re-dispatch market power; Real-time grid operation; Machine Learning; European electricity market
\vspace{0.15in}\\
\noindent\textbf{JEL Codes:} C4; C5; D4; L9; Q4\\
\end{abstract}

\setcounter{page}{0}
\thispagestyle{empty}
\end{titlepage}
\pagebreak \newpage

\doublespacing

\section{Introduction} \label{sec:introduction}
The response of governments around the world to the COVID-19 pandemic has led to negative demand shocks to almost all industries, particularly those in the energy sector. Oil-prices plummeted and the West Texas Intermediate (WTI) futures contract for delivery in May 2020 went negative on April 20 reflecting the exhaustion of local oil storage capacity \citep{Bore20}. Industrial production has halted, shops and offices were closed, and electrified public transport operated at reduced service, all of which reduced the demand for electricity and its pattern across time and space. 

In this paper, we explore the consequences of the particularly strict lockdown in Italy in the spring of 2020 on the performance of the country's wholesale electricity market. The lockdown significantly reduced the demand for controllable sources of electricity such as thermal generation units and hydro units with storage capabilities. These units serve \textit{net demand}---the difference between system demand and supply of non-controllable sources that include renewables such as wind, solar, non-storable hydro, and net imports.\footnote{Because renewables have a close to zero variable cost of producing energy, these resources will be almost always operated when the underlying resource is available. Net-imports are deemed to be firm after the day-ahead market-clearing and are therefore another fixed source of supply for system operators to deal with in the real-time re-dispatch process.  Transmission system operators in Europe do have the ability to change net imports close to real-time but only in extreme situations to solve real-time security issues. New European platforms for trading balancing resources closer to real-time are also currently under consideration.} 

Consequently, the negative COVID-19 electricity demand shock translates into a negative net demand shock because the supply of non-controllable sources were largely unchanged during the lockdown. Therefore, lockdowns and their associated low net demand realizations can provide insight into the challenges system operators may face as regions increase the share of intermittent renewables in their electricity supply industries.  In this sense, the COVID-19 lockdown provides a unique opportunity to analyze potential weaknesses of current electricity market designs with a higher share of intermittent renewables envisioned by the climate policy goals of many countries around the world.\footnote{Because of the intermittency of wind and solar energy production, an increase in wind and solar generation capacity is likely to lead to a more volatile net demand than the equivalent average net demand reduction due to the lockdown demand reduction.}

A back of envelope calculation reveals that the 20\% decrease in business-as-usual (BAU) demand caused by the lockdown in Italy is the equivalent to a 2.3 times higher output from wind and solar energy at pre-COVID-19 demand levels.\footnote{Average hourly demand between March and April over the years 2017 to 2019 was 31.6\;GWh and average hourly generation from wind and solar combined was 4.9\;GWh. A 20\% decrease in average demand (0.2 $\times$ 31.6\;GWh =  6.32\;GWh) is equivalent to an increase of hourly generation from wind and solar by factor 2.3 to 11.22\;GWh ( = 4.9\;GWh + 6.32\; GWh).} More than doubling the output from wind and solar may sound overly ambitious but it is well within the targets for renewable energy production in many countries around the world.

Intermittent renewables such as wind and solar, are likely to concentrate their production within certain hours of the day, month, or year, which can significantly exacerbate the re-dispatch cost increase we identify.\footnote{For example, California produces more than double the amount of wind and solar energy in the summer months relative to other months of the year.} From an environmental perspective, the first-order effect of additional renewable capacity is that emissions will decrease because generation from thermal units will be displaced. However, the intermittent nature of many renewable technologies is likely to increase the importance of a second-order effect that causes a more inefficient operation of remaining thermal generation units because of more start-ups and faster ramps of these units \citep[see e.g.,][]{GrMa17, KaBe20}. The cost of additional start-ups and faster ramps associated with responding to the rapid appearance and disappearance of wind and solar energy can scale rapidly with the amount of renewable energy.\footnote{\cite{ScPa17} estimate that the overall number of start-ups would grow by 81\% (costs by 119\%) for Germany between 2013 and 2030 as the share of variable renewables is expected to grow from 14\% to 34\% if no investments in more flexible technologies including storage are made.}
 
A negative demand shock paired with lower input prices to produce electricity should lead to lower electricity prices. In Figure~\ref{fig:punReDispatch}, Panel (a), we show average hourly day-ahead market electricity prices were down by \deltaPUN during the period of the lockdown compared to BAU levels. However, in simplified electricity market designs that do not account for intra-zonal transmission constraints and other relevant system security constraints in the day-ahead market that exist in virtually all European countries and most wholesale markets outside of the United States, a re-dispatch process is necessary to adjust day-ahead market schedules to ensure that they do not violate real-time transmission network and other system security constraints \citep[see, e.g.,][for more details]{GrQu20, GrQuIncDec20}. Particularly in simplified electricity market designs without an effective local market power mitigation mechanism in place, this re-dispatch process is likely to become more costly as the share of intermittent renewable resources increases because a larger share of the available controllable generation capacity is likely to have to be adjusted in the re-dispatch process to achieve schedules that are compatible with a secure operation of the grid.

In Figure~\ref{fig:punReDispatch}, Panel (b), we show average hourly re-dispatch costs per MWh of demand up by \deltaRedispatchCostDemand relative to the average for the same time period in previous years, what we call the BAU period.\footnote{In absolute terms the re-dispatch costs are up by \deltaRedispatchCost relative to the same time period during previous years. Figure~\ref{fig:reDispatchCost} compares the hourly average re-dispatch costs per week during the lockdown versus the same time during previous years.} While the average BAU period re-dispatch costs per MWh of demand was about 18\% of the average day-ahead market price, it increased to 71\% of the average daily day-ahead market price during the lockdown. Furthermore, in the 20\% highest re-dispatch cost days during the lockdown, the average re-dispatch cost per MWh of demand exceeded the average daily day-ahead market price.

The increase in re-dispatch costs during the lockdown has significantly reduced the cost savings to final consumers due to the day-ahead market price decrease from lower net demand during the lockdown. There are two major explanations for this result: First, this demand shock created additional opportunities, not available to suppliers outside of the lockdown period, to profit from the divergence between the network model used to clear the day-ahead market and network constraints necessary to operate the grid in real-time as discussed in \cite{GrQuIncDec20}.\footnote{In order to ensure a secure operation of the power system, generation units providing ancillary services should be distributed throughout the transmission network. The probability that the schedules that emerge from the day-ahead market meet this requirement decreases when a lower number of power plants are dispatched due to a low net demand. Particularly at low net demand levels, these locational requirements create relatively small local markets with a high concentration of generation ownership, which increases the ability of each single market participant to affect outcomes in these local markets.} Second, this persistently low level of net demand is likely to require additional security constraints to be respected in operating the grid during a larger fraction of hours of the day.

To compute a BAU re-dispatch cost counter-factual that allows us to distinguish between these two determinants of increased re-dispatch costs, we estimate the relationship between hourly re-dispatch costs using historical data on system conditions (including net demand) from January 1, 2017 to December 31, 2019. We use a deep-learning neural network model to predict BAU re-dispatch costs given system conditions during the lockdown period.\footnote{\cite{LaFj18} find that deep-learning approaches outperform traditional regression based time-series forecasting methods to predict hourly electricity prices. Within the class of deep-learning models, they find that a deep neural network with two layers outperformed other deep-learning models in terms of prediction accuracy. \cite{BeAl20} also deploy machine learning methods to study the effect of the pandemic on the French electricity market focusing mainly on day-ahead market performance and consequences of the price drop for market participants.}

We find that predicted BAU hourly re-dispatch costs given system conditions for the lockdown period are only 26\% higher than our BAU period re-dispatch costs. This counter-factual estimate of the increase in re-dispatch costs is approximately one-third of the \deltaRedispatchCost percent increase in the average hourly re-dispatch costs during the lockdown period relative to our BAU period re-dispatch costs. These two results suggest that there are likely to be new offer strategies that suppliers with controllable resources in their portfolio can employ to exercise unilateral market power during the persistently low (net) demand hours that occurred during the lockdown.\footnote{Note that our predictive model estimated over previous years embodies the ability of suppliers to exercise unilateral market power during the periodic low net demand levels that occur on weekends and holidays during this time period. Moreover, this time period also contains a number of low net demand periods of a similar magnitude to those to that occurred during the lockdown period.}  However, we also recognize that some of this re-dispatch cost increase could have been driven by an increased number of operating constraints that must be respected during these persistent low-net demand conditions.

The result that a model estimated using data from 2017--2019 predicts re-dispatch costs during the lockdown period that are a fraction of re-dispatch costs during the lockdown is robust to a variety of different model specifications, including one that attempts to account for dynamic ramping constraints throughout the day faced by controllable thermal resources. We also use our BAU model to estimate how an increase in the amount of renewable energy would affect re-dispatch costs without the lockdown demand reduction. We find that doubling the output from renewable resources would increase re-dispatch costs by 37\% during the pre-lockdown period of January 1, 2017 to March 7, 2020. This result reinforces our conclusion that re-dispatch costs are likely to increase significantly as a result of an increasing share of intermittent renewables at current demand levels.

Although the market response to an unexpected persistent net demand reduction caused by the COVID-19 lockdown is likely to be different from a more gradual net demand reduction caused by increased investments in wind and solar resources, our results demonstrate that without investments in transmission expansions and other technologies for managing low net demand as well as an effective local market power mitigation mechanism, the levels of re-dispatch costs could rise rapidly. At these low net demand levels many system stability constraints bind which can create new opportunities for suppliers providing these services to increase the prices they are paid.

These results also underscore the need for regions with ambitious wind and solar energy goals to adopt wholesale market designs that more closely match the economic model used to set prices and generation output levels to the way the transmission network is actually operated.\footnote{See \cite{GrQuIncDec20} for an example of market participant behavior that can arise from a market design that does not match the economic model used to set prices and output level to the way the system is actually operated.} Our results demonstrate that the opportunities for suppliers to profit from the difference between the model used to operate the electricity market and how the grid is actually operated scales rapidly as the average level of net demand falls.

The remainder of the paper is organized as follows. In Section~\ref{sec:market}, we describe the key features of structure and operation of Italian electricity supply industry necessary to understand our analysis. In Section~\ref{sec:shock}, we show how the lockdown demand shock has affected market outcomes in the Italian electricity market. In Section~\ref{sec:empirics}, we detail our approach to estimating the pre-COVID-19 relationship between system conditions and re-dispatch costs that we subsequently use to predict counterfactual lockdown re-dispatch costs. In Section~\ref{sec:res}, we present our results and investigate their robustness under alternative modeling assumptions. We conclude the paper in Section~\ref{sec:conclusion}.

\section{The Operation of the Italian Wholesale Electricity Market}\label{sec:market}
The Italian wholesale electricity market consists of the European day-ahead market followed by a series of domestic intra-day market sessions, and finally the real-time re-dispatch market. The day-ahead market does not procure ancillary services, only energy. In the intra-day market sessions, participants have the option to update the generation and demand schedules that emerge from the day-ahead market or a previous intra-day market session. The day-ahead market as well as all of the intra-day markets are zonal-pricing markets that ignore transmission network constraints within the zone and other relevant generation unit operating constraints in setting prices and generation unit output levels.\footnote{Currently, the day-ahead market and intra-day markets consist of seven bidding zones (see Tables~\ref{tab:data} and \ref{tab:desc} for more details).}

Shortly after the day-ahead market clears, two out of the seven intra-day market sessions are run, still one day in advance of actual system operation. After the clearing of the second intra-day market, the first session of the re-dispatch market takes place. Five other re-dispatch sessions will be run, one after each intra-day market session as well as a real-time re-dispatch market session that clears every fifteen minutes.

In the real-time re-dispatch market, the objective is to balance any net demand forecast errors but also to transform the schedules resulting from the zonal day-ahead and intra-day market-clearing processes into final schedules that allow secure grid operation in real-time by minimizing the combined as-offered and as-bid cost to change generation schedules. Generation units that are needed to produce more output are paid as-offered to supply this energy and generation units that are unable to produce as much energy because of a real-time operating constraint sell this energy as-bid. The solution to this optimization problem accounts for a nodal network model, the possibility that equipment can fail, errors in forecasts of demand or non-controllable generation, and ensures that technical parameters such as frequency levels or voltage levels are within their security ranges. An offer to start-up a unit or to change a unit's configuration can be submitted to the real-time re-dispatch market as well as price/quantity pairs to increase and decrease a units schedule. The re-dispatch market is operated by the Italian transmission system operator (Terna). Between 2017 and 2019 the average annual real-time upward re-dispatch volume was 16\;TWh and downward re-dispatch volume was 19\;TWh. More details on the market design can be found in \cite{GrQu20, GrQuIncDec20}.

\cite{GrQuIncDec20} find that market participants factor in the expected revenues they can earn from being accepted in the re-dispatch market when they formulate their offers into the day-ahead market. Suppliers recognize that the real-time operating levels of all generation units must respect all network and generation unit-level operating constraints, whether or not these constraints are accounted for in the day-ahead or the intra-day market-clearing engine. Differences between the constraints on generation unit behavior that must be respected in the day-ahead and intra-day markets and the additional constraints that must be respected in the real-time operation of the transmission network are what create the opportunities for suppliers to play what has come to be called the ``INC/DEC Game.''

Ignoring the forecast error in locational net demand profiles between day-ahead and real-time, demand for re-dispatch energy from a generation unit upward or downward arises if a unit's day-ahead market schedule is not compatible with secure operation of the grid in real-time. The ``INC/DEC Game'' relies on the fact that the demand for re-dispatch from a generation unit is endogenously determined by the owner's day-ahead market offer and the day-ahead market offers of other market participants. A high offer price in the day-ahead market can cause a unit required to supply energy in real-time to fail to sell energy in the day-ahead market. A low offer price in the day-ahead market can cause a unit that cannot supply energy in real-time to sell energy in the day-ahead market.

This logic implies that a generation unit owner that is confident its unit is required to run in real-time may offer this unit in the day-ahead market at extremely high price. The unit would either be taken in the day-ahead market at this price or not taken in the day-ahead market but subsequently taken in the re-dispatch market at this offer price or an even higher offer price. The more confident the unit owner is that its unit will be needed to supply energy in real-time regardless of its offer price in the re-dispatch market, the higher the offer price the unit owner can submit into the day-ahead market. 

Similar logic applies to the case of suppliers that are confident that their generation units cannot supply energy in real-time because of a transmission network or other operating constraint.  In this case, the unit owner would submit a very low offer price into the day-ahead market to ensure that it sells energy at the market-clearing zonal price. The more confident the unit owner is that this energy cannot be supplied in real-time, the lower is the offer price submitted into the day-ahead market. In the re-dispatch market this unit owner will then buy back this energy at a bid price that is lower than the market-clearing zonal price and earn the difference between the day-ahead zonal price and this bid price times the amount of energy it is unable to supply.  

In regions that employ zonal day-ahead and intra-day markets and operate a pay-as offered and buy as-bid re-dispatch process, the opportunities for controllable generation units to profit from the predictability of net demand conditions that make their units necessary to operate or not operate are likely to increase as the amount intermittent renewable generation increases.\footnote{Investments in resources that provide flexibility, such as storage, demand response, and transmission network upgrades can reduce the frequency that these opportunities arise.}
In Table~\ref{tab:resCap}, Panel A, we detail the actual installed capacity of wind and solar between 2012 and 2020 in the Italian market. Installed capacities of solar has been steadily increasing from 17\;GW to 21\;GW and of wind from 8\;GW to 11\;GW. In Panel B, we show several projections for the years 2025, 2030, and 2040. Notably, solar capacity is projected to more than double in two out of three scenarios for 2030. Wind capacity is also projected to increase by more than 60\% by 2030 according to the the same two scenarios.

\begin{table}[ht] 
  \centering
  \caption{Actual and Projected Solar and Wind Capacity}
  \begin{threeparttable}
    \begin{tabular}{p{3cm}p{3cm}p{3cm}}
    \toprule
    Year  & Solar & Wind \\
    \midrule
    \multicolumn{3}{c}{\textit{Panel A:} Actual} \\
    2012 & 16.78 & 8.06 \\
    2013 & 18.19 & 8.50 \\
    2014 & 18.59 & 8.67 \\
    2015 & 18.89 & 9.20 \\
    2016 & 19.30 & 9.42 \\
    2017 & 19.68 & 9.78 \\
    2018 & 20.12 & 10.31 \\
    2019 & 20.90 & 10.76 \\
    2020 & 21.20 & 10.81 \\ \addlinespace[5pt]
    \multicolumn{3}{c}{\textit{Panel B:} Projections} \\
    2025\tnote{1} & 26.59 & 15.42 \\ \addlinespace[5pt]
    2030\tnote{1} & 50.00 & 17.52 \\ 
    2030\tnote{2} & 30.48 & 13.61 \\
    2030\tnote{3} & 49.33 & 18.89 \\ \addlinespace[5pt]
    2040\tnote{2} & 47.48 & 17.61 \\
    2040\tnote{3} & 69.83 & 25.38 \\
    \bottomrule
    \end{tabular}%
    \begin{tablenotes}
    \item \tnote{1} Piano Nazionale Integrato per l'Energia e il Clima (National integrated Energy and Climate Plan)
    \item \tnote{2} Business-as-usual projection
    \item \tnote{3} Strong growth in distributed generation projection (Decentralized generation scenario) 
    \item \textit{Notes}: Capacity values in GW. Only solar photovoltaics and wind on-shore considered. Actual capacity values publicly available from \url{https://www.terna.it/en/electric-system/dispatching/renewable-sources} and projected values from \url{https://download.terna.it/terna/Dati%20generazione%20PdS_V3_2020_exact_8d7f5cdeb456feb.xlsx}.
    \end{tablenotes}
    \end{threeparttable}
  \label{tab:resCap}
\end{table}

\section{Negative Lockdown Demand Shock and its Effects}\label{sec:shock}
Large regions around Lombardy in Northern Italy---the economic and industrial powerhouse of the country---shut down on March 8, 2020 and the country-wide lockdown followed suit on March 10, 2020. In the days that followed, the lockdown became even more stringent by narrowing the definition of what an essential business is. The strict lockdown was eased on April 26, 2020.

\begin{figure}[ht]
\caption{Demand-Shock Due to Lockdown in Response to COVID-19 Pandemic}
    \centering
	\subfigure[]{\includegraphics[width=.49 \textwidth]{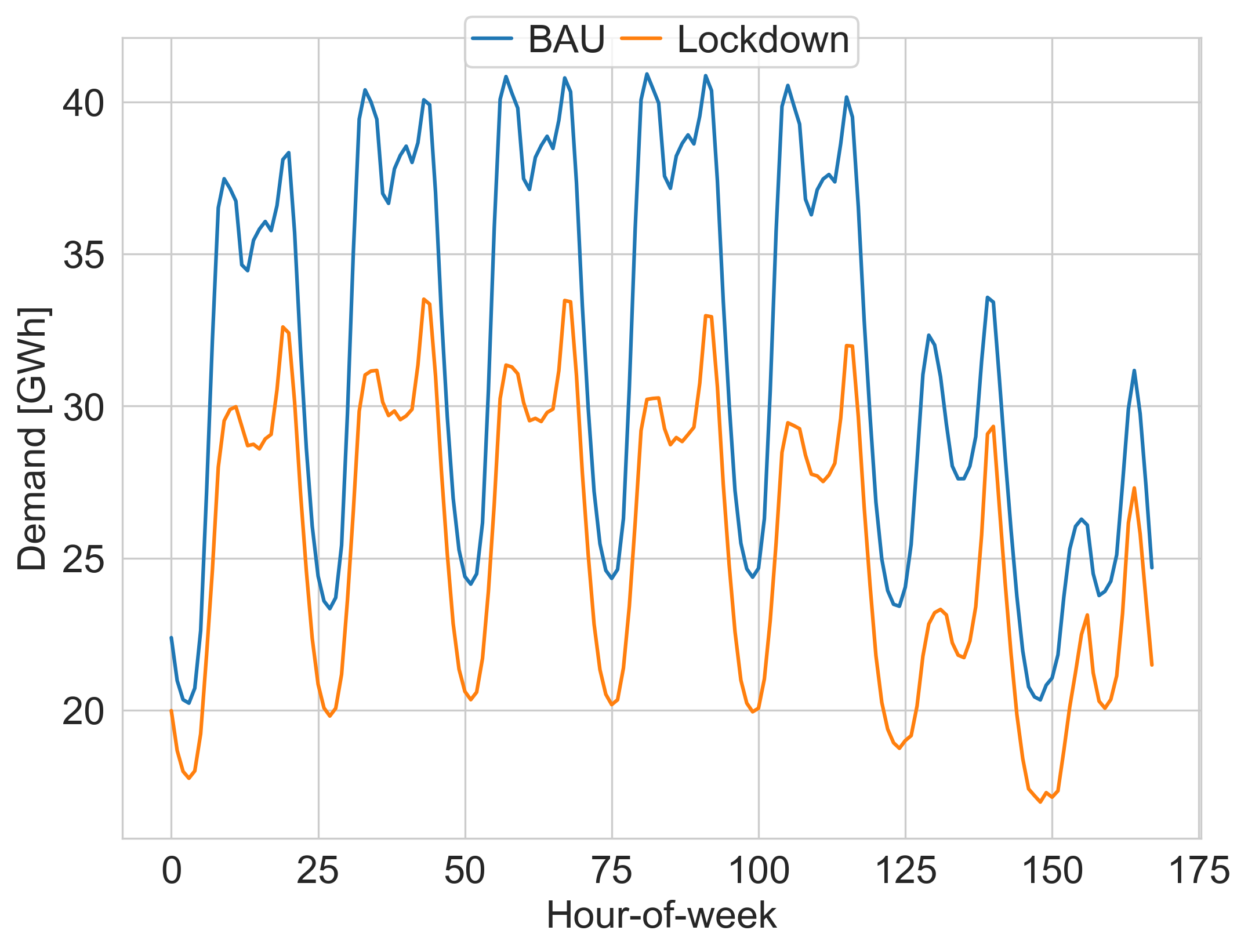}}
    \subfigure[]{\includegraphics[width=.49 \textwidth]{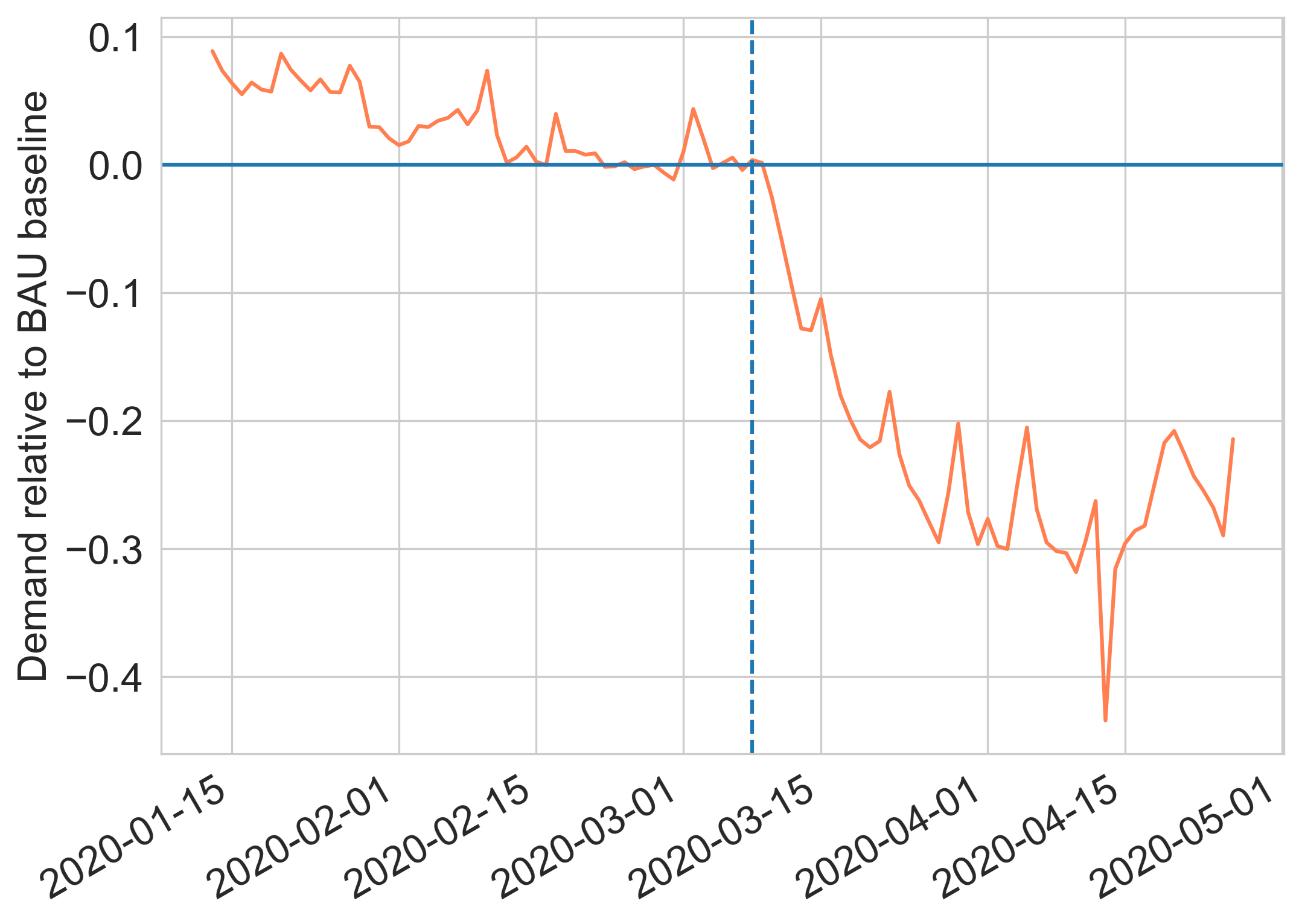}}
    \floatfoot{\footnotesize \textit{Notes:} Panel~(a): Business-as-usual (BAU) demand (market load) calculated as the average demand for each hour of the week over March and April in the years 2017--2019. Lockdown demand calculated as the average demand between March 9, 2020 and April 26, 2020. The first hour of the week starts on Monday 00:00 AM. Panel~(b): Daily demand change relative to daily baseline demand, that is, the daily average demand for each day-of-week between January and April over the years 2017--2019. Dotted vertical line indicates the date of when the lockdown began (March 8, 2020). Data sources to derive the graph described in Table~\ref{tab:data}.}
    \label{fig:shock}
\end{figure}

In Figure~\ref{fig:shock}, we show how the lockdown of effectively all non-essential businesses in a response to the COVID-19 outbreak drastically reduced the national demand for electricity. In Panel (a), we compare the BAU average weekly demand profile, that we define as the hourly average demand in March and April during the years 2017 through 2019 for each hour of the week, to the demand profile during the seven weeks of lockdown (March 9, 2020 until April 26, 2020),\footnote{We define the lockdown period throughout the paper to range between Sunday, March 8, 2020 and Sunday, April 26, 2020. However, for the graphs showing weekly averages, we decided to skip hourly observations from Sunday, March 8, 2020, to obtain the same number of observations for each week.} where the first hour of the week is the hour beginning Monday at 00:00 AM. The figure demonstrates that the average hourly demand is lower in all hours of the week during the lockdown period. In Panel (b), we show how the daily average of demand has changed relative to the daily average demand for each day-of-week between January and April over the years 2017 through 2019. Average daily demand for the lockdown period is \deltaDemand less than the average daily demand during same time period in 2017 through 2019. 
An increasing number of electricity markets have a considerable share of non-controllable supply. Non-controllable supply includes generation from renewables, such as wind, solar, or hydro.  In Europe, net-imports made through either long-term allocation processes or the joint European day-ahead market-clearing are non-controllable in real time except for emergency situations. Higher net imports and more output from non-controllable units will reduce the demand that will be served by controllable generation units, including fossil-fuel generators but also storage units. 

We define the hourly net demand ($ND$) for controllable generation units for each bidding zone $z$ as follows

\begin{equation}\label{eq:rd}
    ND_z = D_z - RES_z - Hydro_z - Imp_z,
\end{equation}

\noindent
whereas $D$ is system demand, $RES$ the supply from intermittent renewable sources such as wind or solar, $Hydro$ is non-dispatchable hydro production, such as generation from run-of-river plants, and $Imp$ is net imports from foreign countries. We use the same logic to create day-ahead net demand \textit{forecast} ($ND^{FC}$) variables by replacing $D$ and $RES$ with their day-ahead forecasts. Note that non-controllable hydro schedules as well as net imports from foreign countries are firm after the day-ahead market-clearing.

Figure~\ref{fig:shockRd} shows hourly net demand boxplots for the BAU period and the lockdown periods. BAU period net demands are from hours in March to April for the years 2017 through 2019. Lockdown net demands are for hours between 2020-03-08 and 2020-04-26. Boxes represent the interquartile range (IQR) and the upper and lower vertical bars are the 1 percentile and 99 percentile for that hour of the day. Diamonds represent outliers not included in the 1 to 99 percentile range. The average hourly lockdown period net demand was 21\% lower than the BAU period average hourly net demand.

\begin{figure}[ht]
\caption{National Net Demand Duration Curves}
    \centering
	\subfigure[]{\includegraphics[width=.49 \textwidth]{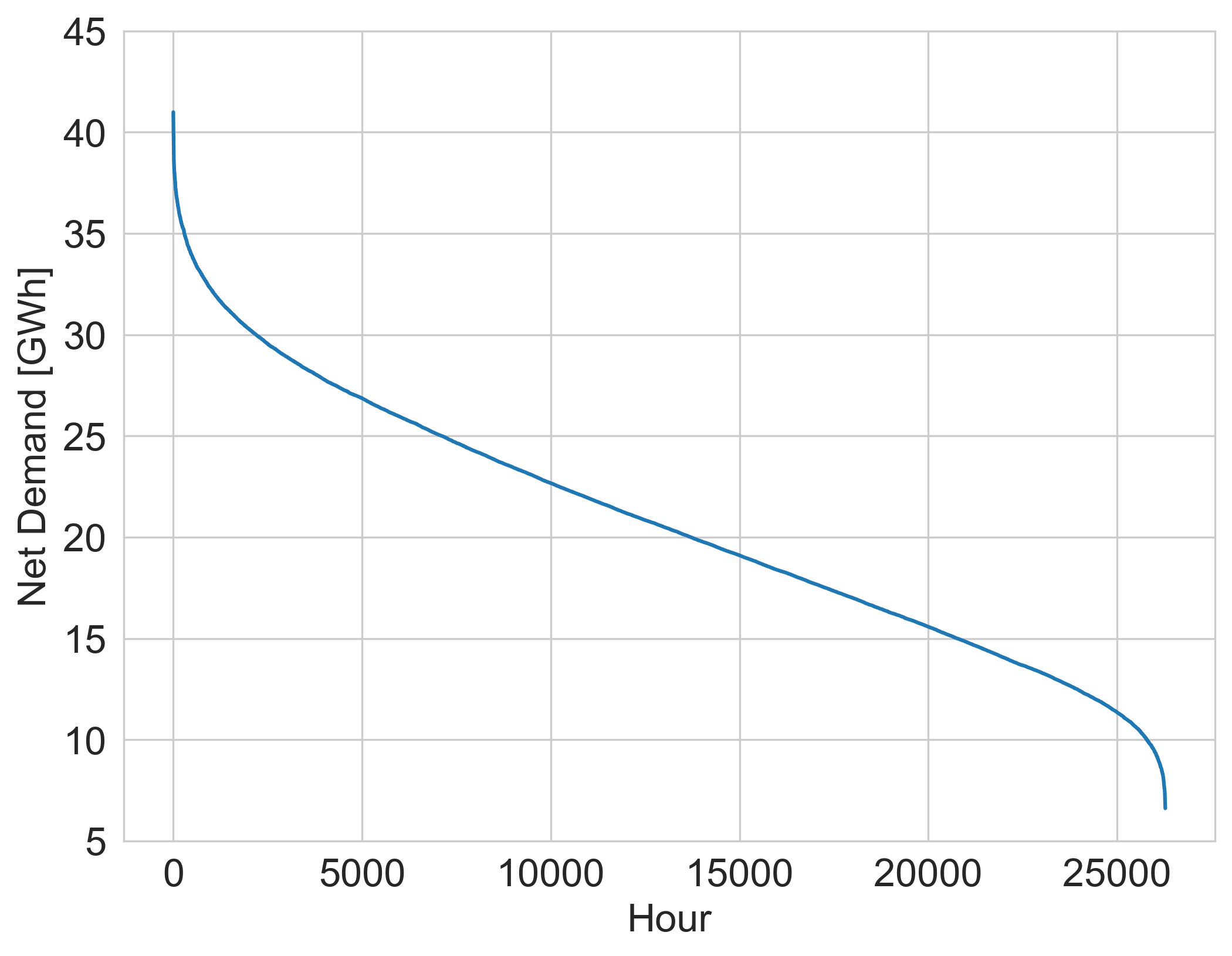}}
    \subfigure[]{\includegraphics[width=.49 \textwidth]{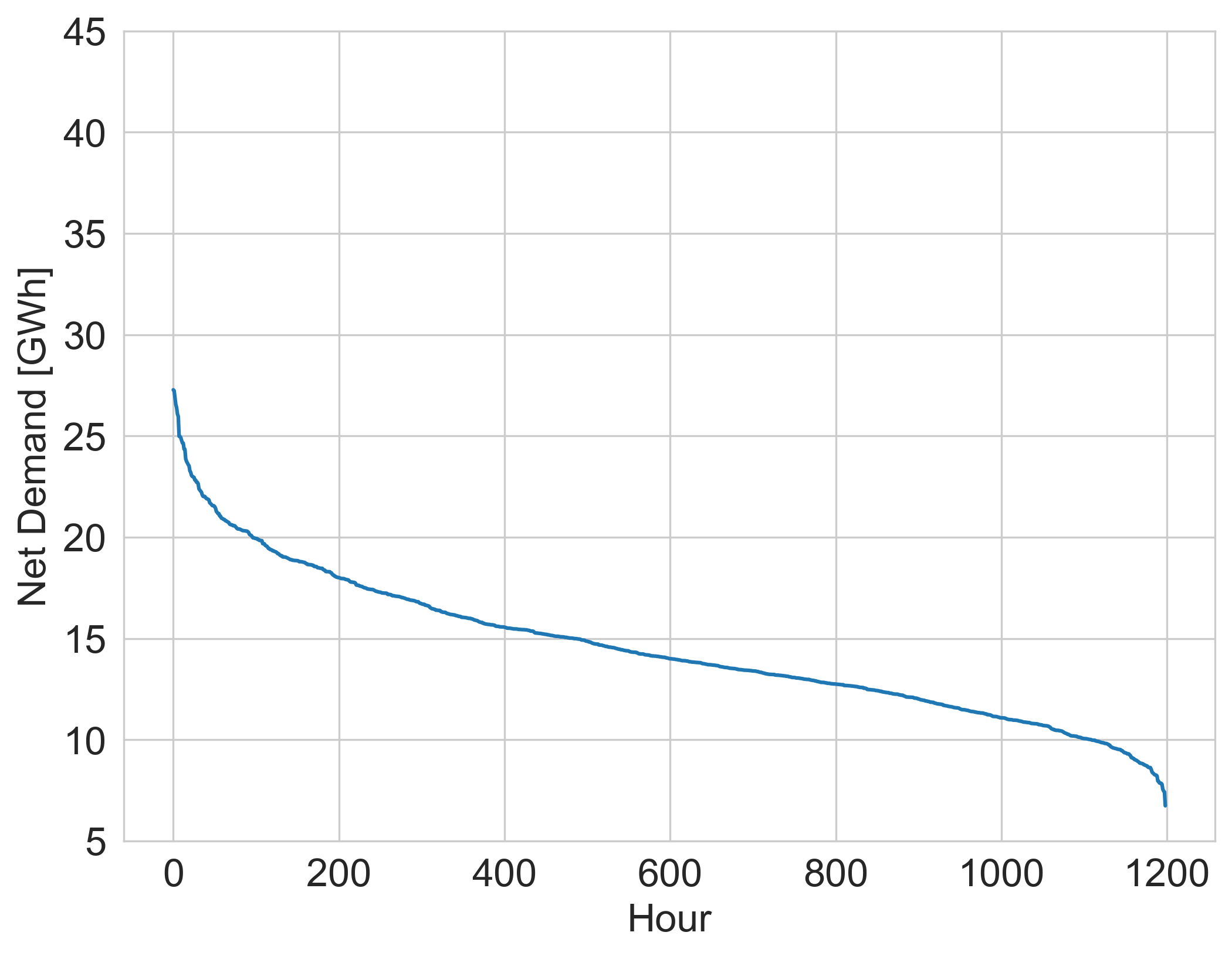}}
    \floatfoot{\footnotesize \textit{Notes:} Panel~(a): Hours in the years 2017--2019. Panel~(b): Hours during lockdown period (2020-03-08 to 2020-04-26).}
    \label{fig:netLoadDur}
\end{figure}

Panel (a) of Figure \ref{fig:netLoadDur} plots the net demand duration curve for the years 2017 through 2019. Panel (b) plots the net demand duration curve for the lockdown period from 2020-03-08 to 2020-04-26. Although the range of net demands from 2017 to 2019 is larger than the range of net demands during the lockdown period, the range of net demand from 2017 to 2019 contains the range of net demands during the lockdown period. The shape of the net demand duration curve in panel (a) is similar to the shape of the net demand duration curve in panel (b). The major difference between the two curves is that much more probability mass is concentrated in a much smaller range of low demand levels during the lockdown period.

\begin{figure}[ht]
\caption{Electricity Prices versus\ Re-dispatch Costs}
    \centering
	\subfigure[]{\includegraphics[width=.49 \textwidth]{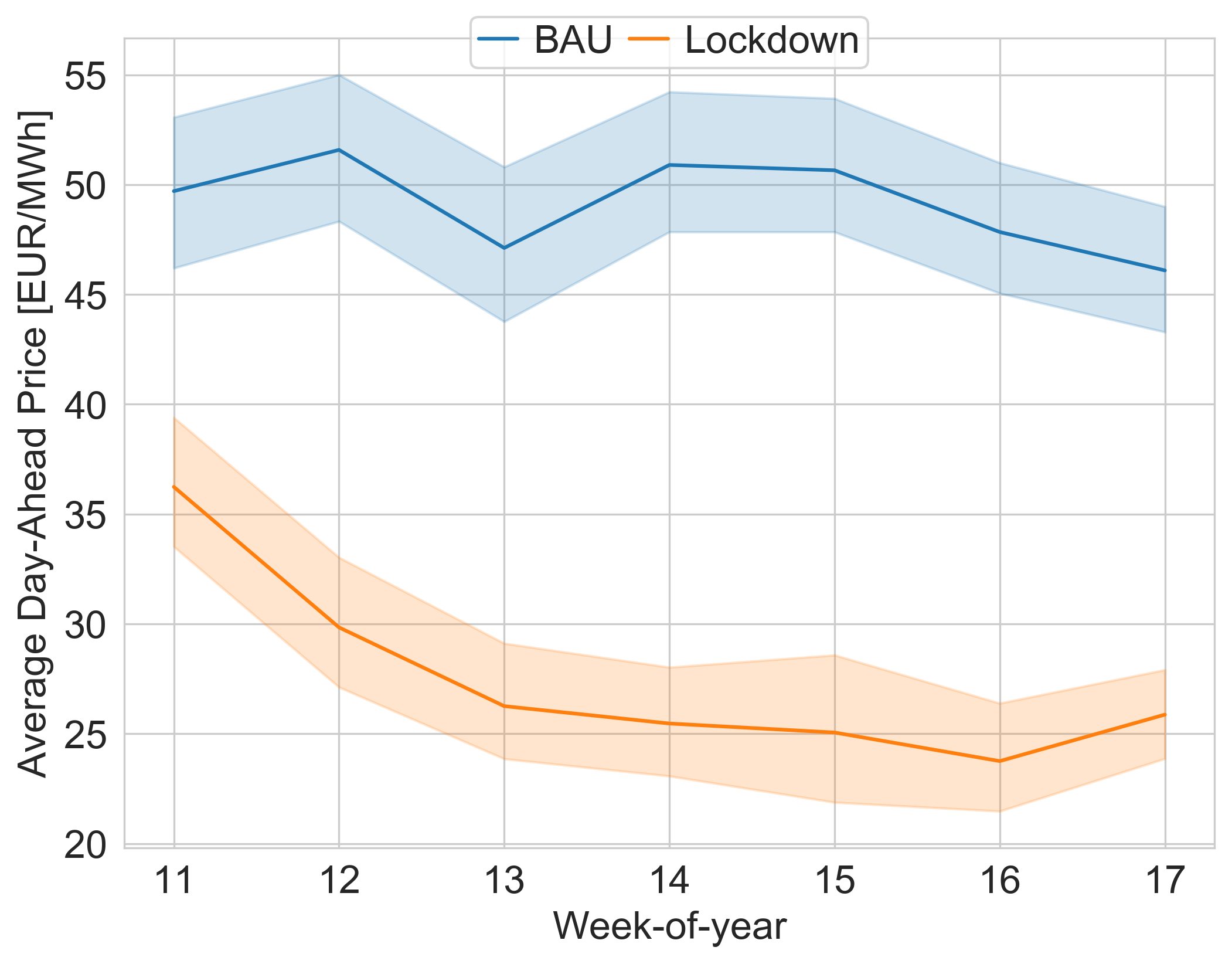}}
    \subfigure[]{\includegraphics[width=.49 \textwidth]{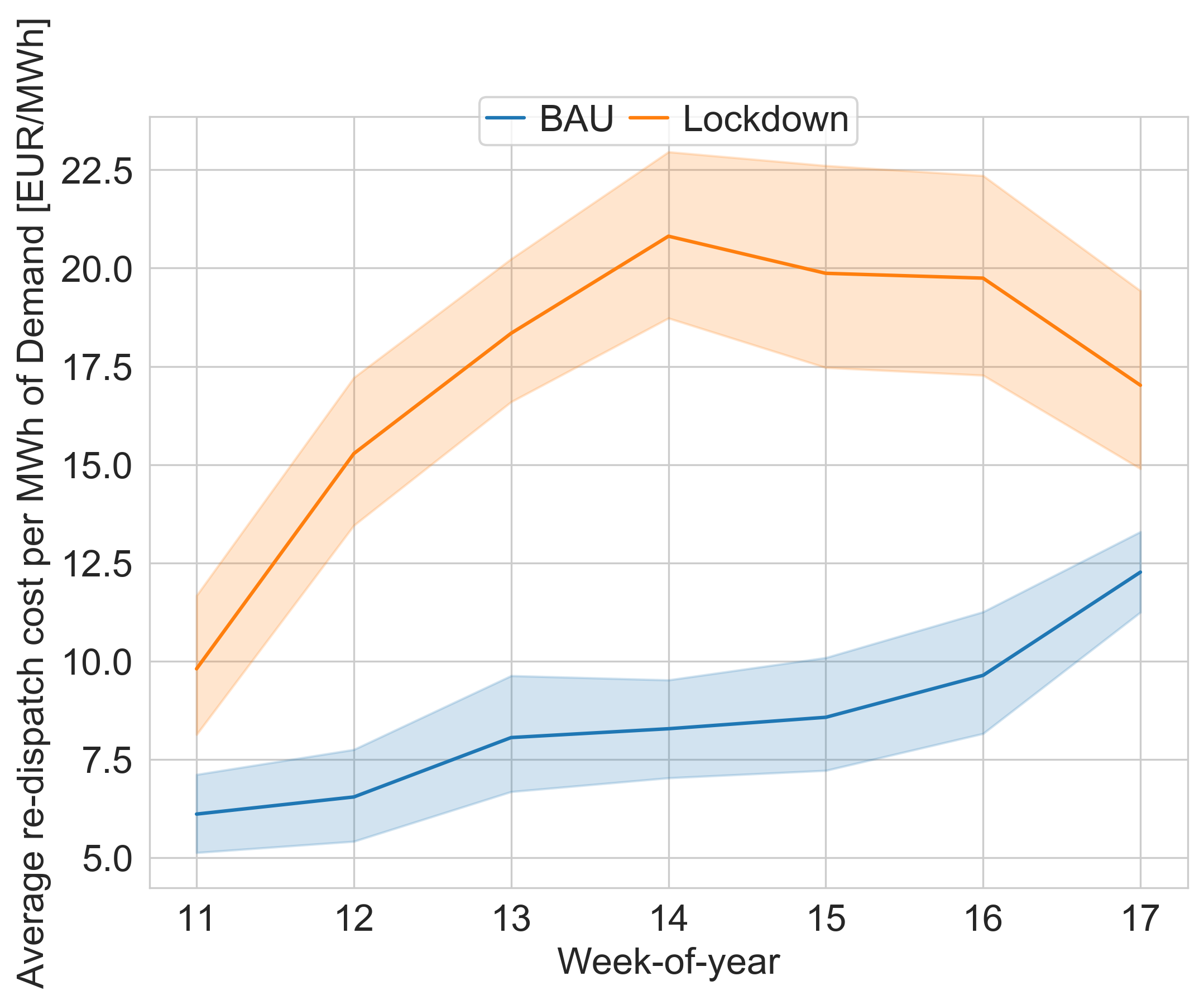}}
    \floatfoot{\footnotesize \textit{Notes:} Panel~(a): Business-as-usual (BAU) day-ahead market electricity price calculated as the average price for each hour of the week over March and April in the years 2017--2019. Lockdown electricity price calculated as the average price between March 9, 2020 and April 26, 2020. Panel~(b): BAU re-dispatch costs calculated as the average costs per MWh of demand for each hour of the week over March and April during the years 2017--2019. Lockdown re-dispatch costs calculated as the average hourly costs per MWh of demand between March 9, 2020 and April 26, 2020. Shaded area around the mean values represents the pointwise 95\% confidence interval. Data sources to derive the graph described in Table~\ref{tab:data}.}
    \label{fig:punReDispatch}
\end{figure}

In Figure~\ref{fig:punReDispatch}, Panel (a), we show how the negative demand shock affected day-ahead electricity prices compared to BAU. As expected, a negative demand shock paired with lower input prices to produce electricity led to lower electricity prices. Average lockdown period hourly day-ahead market prices were down by 45\% compared to average BAU period hourly day-ahead market prices. Although these lower day-ahead market prices are good news for the final consumer, in simplified European electricity market designs where system security constraints are only accounted for in the real-time re-dispatch market, the final consumer also pays the cost of re-dispatching generation units to achieve physically feasible generation output levels to meet real-time demand at all locations in the transmission network.

In Figure~\ref{fig:punReDispatch}, Panel (b), we show average hourly re-dispatch costs per MWh of demand during lockdown and BAU period applying the same definition for BAU as above.\footnote{Hourly real-time re-dispatch costs are computed as the sum of the awarded incremental real-time re-dispatch quantities valued at the as-offered costs net of the sum of the awarded decremental real-time re-dispatch quantities valued at the as as-bid costs. As-offered costs to start-up a unit or to change a unit's configuration are neglected.} We find that average hourly re-dispatch costs per MWh of demand increased by \deltaRedispatchCostDemand during the lockdown compared to the BAU period. For some weeks during the lockdown, the average re-dispatch cost per MWh of demand was almost as high as the day-ahead market price. The sharp increase in the re-dispatch costs drastically reduces the electricity cost savings for the final consumers due to the reduced electricity demand during the lockdown. 

\section{Estimating BAU Period Re-Dispatch Cost}\label{sec:empirics}
In this section, we estimate a model that predicts the BAU period re-dispatch costs using data on system conditions. This model is then used to predict re-dispatch costs during the lockdown period. Because this model is estimated using data from 2017--2019, the lockdown period predictions from this model cannot capture changes in offer behavior or system security requirements caused by the persistent low net demand levels that occurred during the lockdown period. This model can only use relationship between system conditions and re-dispatch costs during weekends and holidays during the BAU period to predict re-dispatch coss for comparable low net demand periods during the lockdown period.\footnote{In fact, the lowest level of net demand found in our sample was observed in a pre-covid hour, i.e., on a Sunday afternoon in June (2019-06-02, 2 pm to 3pm).}

The main factors used to predict re-dispatch costs are the zonal net demands---the demand within each zone that must to be served by controllable generation units. We also include the day-ahead forecast of zonal net demands. It is important to include actual and forecast zonal net demands because the re-dispatch quantities for each generation unit is the difference between its real-time output and its day-ahead market schedule. For example if demand was underestimated day-ahead, the demand for re-dispatch will be higher because of the forecast error while it will be lower or even negative for an overestimation of day-ahead demand. However, the uncertain forecast error is not the only reason for a (locational) demand for re-dispatch. System operation constraints such as voltage regulation, reserve requirements or nodal network constraints also drive the demand for re-dispatch actions. We include zonal net demand levels because the spatial distribution of net demand reveals (i) inter-zonal power flows\footnote{Across-zone power flows are implicitly defined by zonal balances in a radial network and the current Italian network configuration supports this assumption to a large extent. The zonal net demands result in so-called base-flows and the market power that includes the power to congest a network by market participants with controllable units in their portfolio will affect the final flows through their offer behavior \citep[see, e.g.,][for more details on unilateral market power that includes congestion power in locational markets]{GrWo20}.} and (ii) are useful predictors for whether system constraints will bind. We also include the day-ahead market price to control for the variable cost of the marginal generation unit. We include a workday indicator variable that is equal to one for days that are not weekend days nor holidays. Finally, we include an indicator for the months December to April because the thermal capacity of overhead transmission lines is higher during these months because of lower ambient temperatures levels. Note that our goal is to predict hourly BAU re-dispatch costs for the current day given current system conditions and not to predict re-dispatch costs for a day in the future. Therefore, we believe it is legitimate to use data from the previous day and current day for this purpose.

We compiled data on the variables described above for the time period 2017-01-01 to 2020-04-26. A more detailed description of the variables and their sources can be found in Table~\ref{tab:data}. Descriptive statistics of relevant variables are presented in Table~\ref{tab:desc}.

Instead of using regression based models, we use a deep neural network model with two hidden layers.\footnote{A general introduction to deep learning methods can be found in e.g., \cite{Good16} and \cite{hastie2009elements}} Linear regression models are appealing for constructing counterfactual predictions because they can be fit efficiently using closed form solutions that requires only matrix inversion to derive parameter estimates $\hat{\beta} = (X'X)^{-1} X'y$, with $X$ being the matrix of explanatory variables and $y$ the dependent variable with $y_t$ the $t^{th}$ element of $y$ and $X_t '$ the $t^{th}$ row of $X$. However, obtaining a flexible model to predict out-of-sample values of $y$ may require expanding the set regressors to include polynomials in the elements of $X_t$. However, increasing the number of elements in the vector of regressors is likely to increase the variance of our out-of-sample predictions of $y$ because a number of the true values of these coefficients are likely to be zero, although exactly which ones are is unknown.

The benefit of a deep neural network, however, is its ability to flexibly approximate the conditional mean of $y_t$ given $X_t$ using nonlinear functions of the elements of $X_t$.\footnote{The universal approximation theorem states that a feed-forward network with a single hidden layer containing a finite number of neurons can approximate arbitrary well real-valued continuous functions \citep[see e.g.,][]{Cybe89, Horn91}.} This ability may be very relevant in our context given the complex relationship between re-dispatch costs and input variables. Consequently, if the researcher is willing to tolerate some bias in the estimation of the conditional expectation of $y_t$ given $X_t$, an out-of-sample prediction of $y_s$ given $X_s$ that has smaller expected mean-squared error prediction is possible. A deep neural network assumes $\E(y_t | X_t ) = f(X_t; \beta)$. The parameter vector $\beta$---called weights in the machine learning jargon---of this nonlinear function are chosen to minimize the in-sample mean-squared prediction error as well as accounting for the possibility of within-sample over-fitting. In a two hidden layer neural network three functions are connected in a chain to form $f(x) = f^{(3)}(f^{(2)}(f^{(1)}(x)))$, with $f^{(1)}$ being the first hidden layer, $f^{(2)}$ the second hidden layer, and $f^{(3)}$ being the last layer, called the output layer. The learning algorithm's objective is to optimally use the layers to best approximate $\E( y_t | X_t )$. We refer to Chapter 6 in \cite{Good16}, or Chapter 11 in \cite{hastie2009elements} for more details. The cost of this deep neural network approach is that many model parameters have to be estimated in an iterative process using numerical optimization methods, often with objective function penalties on certain tuning parameters. We detail the model's configuration and our strategies to circumvent known potential shortcomings such as model uncertainty or over-fitting in Appendix~\ref{app:dl}.

A major concern of all machine learning models is over-fitting. We address this issue by using the method of cross-validation for model selection. The basic idea of this method is to split the in-sample data ranging from January 1, 2017 to December 31, 2019 into a training sample and a validation sample. The parameters are estimated on the training sample and the validation sample is used to monitor out-of-training-sample performance (the mean squared error, i.e., the average squared difference between the estimated re-dispatch cost and the actual re-dispatch cost) and set the values of various tuning parameters in estimation samples. The performance on the validation set is used as a proxy for the generalization error and model selection is carried out using this measure see discussed in \cite{Rasm06}. We use a random 70:30 split, i.e., 70\% of the data will be used for training the model and 30\% for validating the model. This approach will make the trained weights (parameter estimates) more generally applicable rather than being too strongly tailored to the estimation or training data. We also set aside a considerable part of the overall data that is not presented to the algorithm for training or for validation (see Appendix~\ref{app:resDes} for more details). We divide this out-of-sample data into ``pre-lockdown''-data ranging from January 1, 2020 to March 7, 2020 and ``lockdown''-data ranging from March 8, 2020 to April 26, 2020 (see Figure~\ref{fig:resDesign} for a graphical summary of the design).

\section{Results}\label{sec:res}
\begin{figure}[ht]
\caption{Distribution Out-of-Sample Prediction Errors}
    \includegraphics[width=.8 \textwidth]{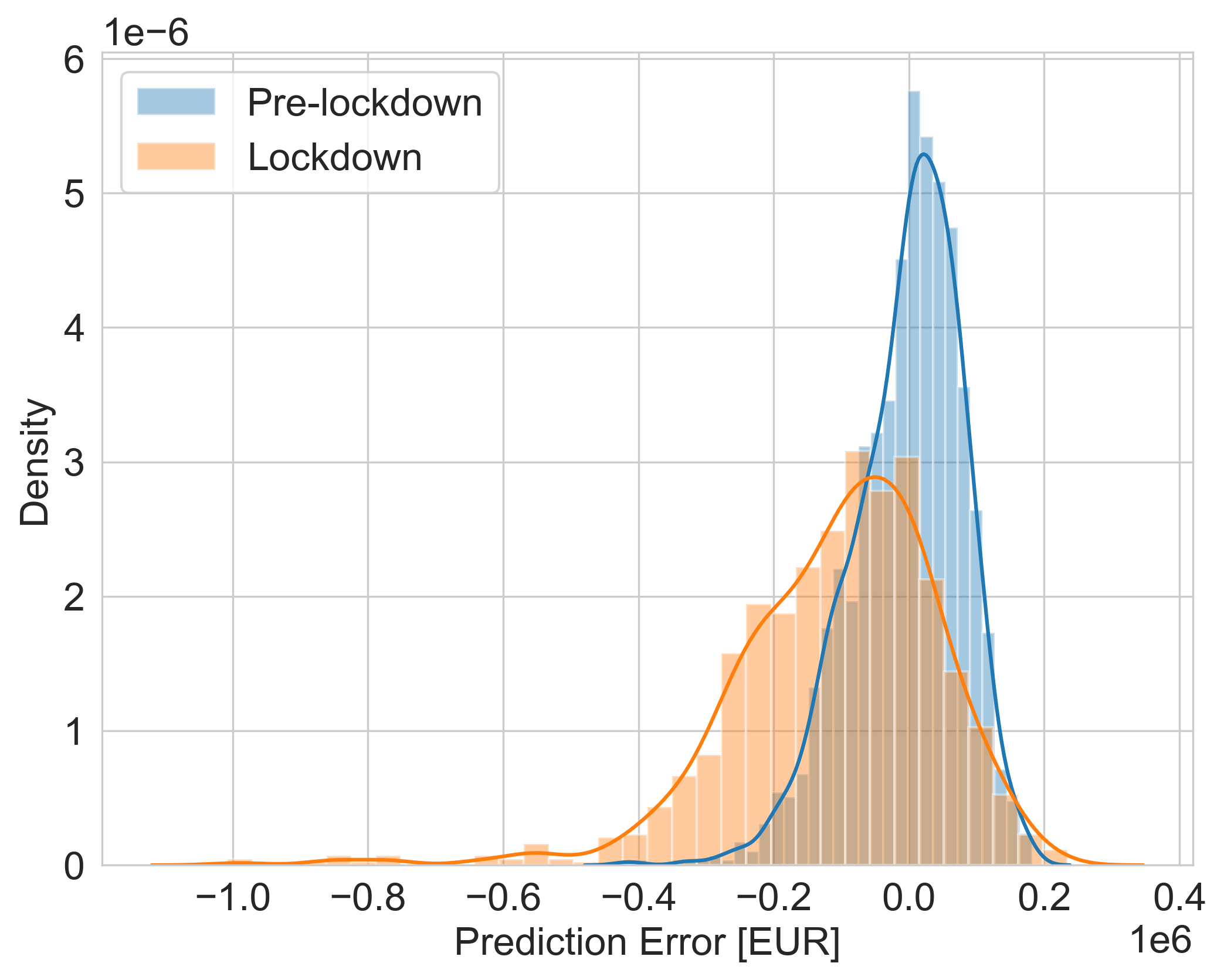}
    \floatfoot{\footnotesize \textit{Notes:} The prediction error is defined as the difference between the predicted out-of-sample pre-lockdown re-dispatch costs and the actual re-dispatch costs realizations. Out-of-sample pre-lockdown period ranges from January 1, 2020 to March 7, 2020 and out-of-sample lockdown period from March 8, 2020 to April 26, 2020.}
    \label{fig:predictionError}
\end{figure}

In Figure~\ref{fig:predictionError}, we present the out-of-sample prediction error for the pre-lockdown period (2020-01-01 to 2020-03-07) and the lockdown period (2020-03-08 to 2020-04-26). The prediction error is defined as the difference between predicted BAU re-dispatch cost from our neural network model and actual re-dispatch cost. Hence a negative prediction error means that we have underestimated the actual re-dispatch cost. The average hourly prediction error during the BAU period is about $-2,000$\;EUR (a 1\% percentage error relative to the average predicted BAU re-dispatch cost) while the average prediction error during lockdown period is orders of magnitude larger at $-107,000$\;EUR (a 37\% percentage error relative to the average predicted BAU re-dispatch cost. Furthermore, the prediction error distribution during the lockdown period is more negatively skewed than the prediction error distribution during the BAU period. A Wilcoxon Signed Rank test comparing the distributions of predicted versus actual re-dispatch costs for the BAU period finds that these distributions are not statistically different at the 5\% level (p-value: 0.23).\footnote{To derive the p-value an asymptotic normal approximation to the null distribution of the test statistic is used.} Applying the same test to the lockdown period yields a p-value that is effectively zero, indicating that the distribution of predicted re-dispatch costs during the lockdown period is statistically significantly different from the distribution of actual re-dispatch costs during this same time period.

\begin{figure}[ht]
\caption{Out-of-sample Predicted Re-dispatch Cost and Actual Realizations}
    \centering
	\subfigure[]{\includegraphics[width=.49 \textwidth]{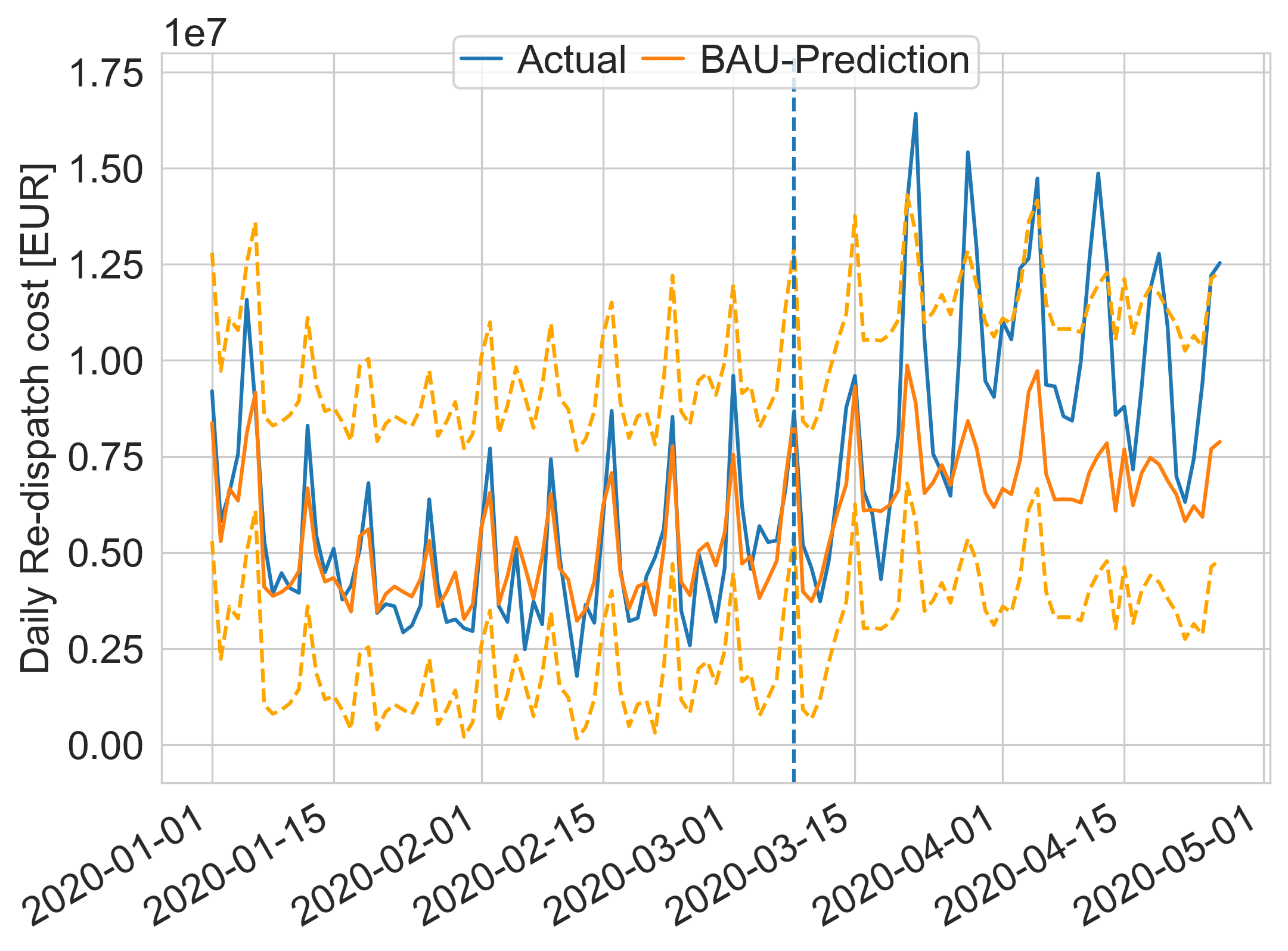}}
    \subfigure[]{\includegraphics[width=.49 \textwidth]{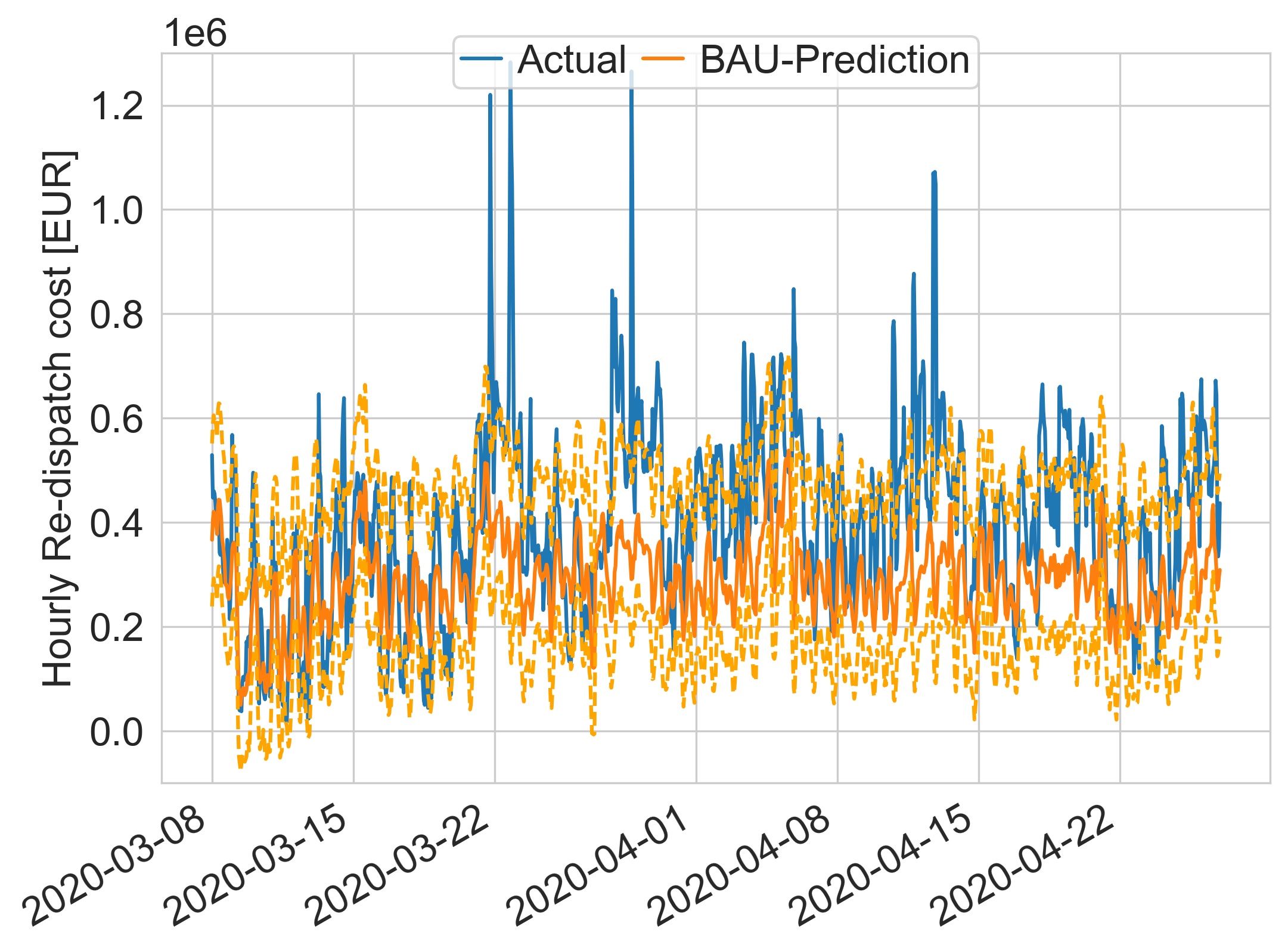}}
    \floatfoot{\footnotesize \textit{Notes:} Panel~(a): Actual daily re-dispatch costs and business-as-usual (BAU) predictions for the out-of-sample period. Dotted vertical line indicates the date of when the lockdown began (March 8, 2020). Panel~(b): Actual hourly re-dispatch costs and BAU predictions during the lockdown. Dotted lines indicate the upper and lower prediction error intervals.}
    \label{fig:prediction}
\end{figure}

In Figure~\ref{fig:prediction}, Panel (a), we compare the daily re-dispatch costs to our predicted BAU re-dispatch costs. We add a prediction error band around our point estimates to account for the uncertainty in the predictions. More precisely, we add the absolute value of the 0.025 quantile of the prediction error during the out-of-sample pre-lockdown period to the point estimates of the lockdown period and subtract the 0.975 quantile from the point estimates. The figure in Panel (a) shows that before the lockdown our model estimates are well within the prediction bands, whereas that is not the case during the lockdown period. In Panel (b), we zoom into the lockdown period and compare hourly values of the actual re-dispatch costs and our predicted BAU re-dispatch costs, showing that realizations in some hours are substantially larger than our predictions.

Overall, we find that the actual average hourly re-dispatch costs during the lockdown were 37\% higher than our BAU estimates during the lockdown and our BAU estimates during the lockdown are 26\% higher than the average hourly re-dispatch costs during the same time period in previous years. As noted earlier, the average hourly re-dispatch cost during the lockdown is \deltaRedispatchCost higher than the average hourly re-dispatch costs during the same time period in previous years. Therefore, roughly two-thirds of the increase in re-dispatch costs during the lockdown can be attributed to the persistent low net demand conditions giving market participants more opportunities to figure out schedule configurations that increase the demand for re-dispatch actions. Furthermore, a higher demand for storage units is created\footnote{Low net demand hours may require to dispatch storage units to increase net demand for accommodating system-relevant thermal generation units. The reservoir balance constraints of storage units make it necessary to also release the stored electricity (in form of water) within a short period of time to have spare head-room in the reservoirs to increase net demand again in the near future if necessary. Therefore, part of the available storage capacity may have considerable market power in settings with persistently low net demand because they face little competition in the market for increasing net-demand in real-time.} and amplified the impact of market power exercise, that cannot be explained by the past relationship between system conditions and re-dispatch costs.\footnote{Although average re-dispatch costs on non-business days before the lockdown are comparable to those on workdays during the lockdown, average hourly re-dispatch costs were 308,000\;EUR for non-business days before the lockdown and 332,000\;EUR only on workdays during the lockdown. This yields a 8\% increase despite lower fuel costs during the lockdown period. The average hourly re-dispatch cost during lockdown using \textit{all} hours yields 395,000\;EUR. Hence, an increase of over 25\% compared to non-business days in March and April over the years 2017--2019.}

The 37\% increase in actual re-dispatch cost compared to our BAU predictions during the lockdown amounts to 129 million EUR for the seven weeks of strict lockdown. In a world with a large share of renewables, the reduced net demand situation would be permanent. Hence, to put things into perspective and extrapolating this amount to an annual level yields an increase in the re-dispatch cost by almost 1 billion EUR per year. As noted earlier, the negative COVID-19 demand led to a net demand reduction that is the equivalent of doubling renewable energy production in Italy.  Consequently, using these numbers implies the potential for a roughly 1 billion EUR increase re-dispatch costs associated with a roughly doubling of renewable energy production in Italy under the existing market design.

\subsection{Robustness Checks}\label{sec:robustness}
In this section, we perform robustness checks of our preferred BAU prediction model. In Section~\ref{Regression_based}, we compare the performance of our neural network model to three regression-based models for computing our counterfactual lockdown period re-dispatch costs and predict lower lockdown period re-dispatch costs than our neural network model. In Section~\ref{sec:res_trend}, we restrict the training and validation samples of our the neural network to 2019 data only to exclude the impact of any pre-existing trends in the share of intermittent renewables in our full 2017 to 2019 sample period. We find even larger counterfactual lockdown period re-dispatch costs using this prediction model compared to our full training and validation sample model. 

In Section~\ref{sec:res_dynModel}, we include dynamic effects---lags of hourly system net demand as well as leads of hourly system net demand forecast---in our neural network model to predict re-dispatch costs because the operation of non-controllable generation units throughout the day typically depends on both the actual and forecast pattern of net demand throughout the day. These models predict higher counterfactual lockdown period re-dispatch costs than our preferred model, but the difference between the distribution of  predicted re-dispatch costs during the lockdown period from this model and the actual distribution of re-dispatch costs during the lockdown period are still statistically different. 
In Section~\ref{sec:renew_and_redis}, we use a neural network model estimated over our 2017 to 2019 period to predict the impact of doubling of actual renewable output during 2017-01-01 to 2020-03-07. We find that predicted re-dispatch costs during this time period are roughly 37\% above actual re-dispatch costs during this time period. This predicted increase in re-dispatch costs (due a doubling of renewable output) relative to actual re-dispatch costs during the pre-lockdown period is larger than the predicted increase in re-dispatch costs during the lockdown compared to BAU period re-dispatch costs, i.e., average hourly re-dispatch costs during the same time period in previous years. 

\subsubsection{Regression-based Models}\label{Regression_based}
In Table~\ref{tab:perf}, Panel A, we present the pre-lockdown out-of-sample performance of several regression-based models estimated over our 2017 to 2019 training and validation sample period. Unlike in the deep-learning framework, we run the linear regressions on the combined training and validation samples. We use the root mean square error (RMSE) metric\footnote{$\text{RMSE}(y,\hat{y}) = \sqrt{T^{-1} \sum_{t=1}^T (y_t - \hat{y}_t)^2}.$} evaluated at pre-lockdown out-of sample observations and predictions to compare the performance of all our models. Our preferred deep neural network specification described in Section~\ref{sec:res} yields an RMSE of 81,676\;EUR for the period 2020-01-01 to 2020-03-07 (the actual average hourly re-dispatch cost for the same period yields 206,000\;EUR) . Using the same set of explanatory variables $X = \begin{bmatrix} X^C & X^I \end{bmatrix}$ with $X^C$ containing the continuous variables (day-ahead market price, zonal net demands, and zonal day-ahead net demand forecasts), and $X^I$ the indicator variables (workday indicator variable and winter indicator variable), as input in estimating a linear model by ordinary least squares yields an RMSE of 103,120\;EUR. 

As discussed in Section~\ref{sec:empirics}, the advantage of the deep-learning framework is that it specifies a flexible nonlinear function for $\E(y_t | X_t )$. A popular approach to account for more flexible functional forms in regression based model is to include polynomial terms and interaction terms. We therefore include 2$^\text{nd}$ degree polynomials and interaction terms of the continuous explanatory variables. This modification of the regression model pushes the pre-lockdown out-of-sample RMSE down to 85,904\;EUR---a drastic improvement that is nonetheless slightly higher than the goodness of fit of our deep neural network model. A tempting strategy to reduce the pre-lockdown RMSE would be to add higher degree polynomials and interaction terms. However, adding 3$^\text{rd}$ degree polynomials and interaction terms of the columns of $X^C$ to the regression model actually drastically increases the pre-lockdown out-of-sample RMSE to 104,505\;EUR. This is a typical of out-of-sample prediction behavior of an over-fitted linear regression model which can be mitigated by the use of our preferred neural network approach.

All three regression models summarized in Table~\ref{tab:perf} yield lower total predicted BAU re-dispatch costs during the lockdown than our neural network model. Therefore the actual average re-dispatch costs relative to the average predicted BAU re-dispatch costs during the lockdown are larger (+55\% to +80\% compared to +37\% of our preferred deep neural network model). In that sense, our preferred deep neural network model leads to a more conservative conclusion on the additional re-dispatch cost increase during the lockdown compared to the regression based model predictions.

\subsubsection{Trend in In-sample Data}\label{sec:res_trend}
On an annual basis we find that re-dispatch costs have increased in 2019 to 1.83 billion EUR compared to 1.53 billion in 2017 and 1.57 billion in 2018. A potential explanation could be that the output from wind was higher in 2019; 19.9\;TWh compared to 17.5\;TWh in 2017 and 17.3\;TWh in 2018. Wind capacity is mainly concentrated in the South of the country where the transmission network is less extensive, which may explain this difference. To check whether this trend in renewable output over our 2017 to 2019 training and validation sample period could explain the level of our predicted re-dispatch cost increase during the lockdown period, we train our model using only data from 2019. More precisely, we adapt the same cross-validation strategy to circumvent over-fitting as before with the only difference that we perform the random sample split on observations that occurred in the year 2019. All other choices of how we implemented our preferred specification of the neural network including hyper-parameter optimization are unchanged and so are the input variables. Instead of a 37\% additional increase in the actual lockdown re-dispatch costs relative to the predicted re-dispatch costs, we find a 43\% increase when training our prediction model on this reduced sample. 

\subsubsection{Dynamic Effects}\label{sec:res_dynModel}
Although, the day-ahead market is cleared hour by hour, market participants' offer strategies that determine the schedules of their conventional dispatchable units are likely to condition on the recent past and forecasts of near future states of the system because of non-convexities in the units' production functions.\footnote{They non-convexities include start-up and minimimum operating level costs and minimum up-time, minimum down-time, and ramping constraints.} Hence, we add 24 hours of lagged system net demand variables to the set of explanatory variables.\footnote{The regressor matrix in this specification is defined as $X = \begin{bmatrix} X^C & X^I & \left\{ \text{L}^k \left( \sum_z ND_z \right) \right\}_{k\in \left\{1, \ldots, 24 \right\}} \end{bmatrix},$ with L being the lag operator.} This modeling choice slightly decreases the RMSE from 81,676\;EUR to 80,863\; EUR as presented in Table~\ref{tab:perf}, Panel B. Adding 24 hour of leads of the system net demand forecast\footnote{The regressor matrix in this specification is defined as $X = \begin{bmatrix} X^C & X^I & \left\{ \text{L}^k \left( \sum_z ND_z \right) \right\}_{k\in \left\{1, \ldots, 24 \right\}} & \left\{ \text{L}^{-k} \left( \sum_z ND^{FC}_z \right) \right\}_{k\in \left\{1, \ldots, 24 \right\}} \end{bmatrix}.$} slightly decreases the pre-lockdown out-of sample RMSE to 80,363\;EUR. If we were to choose a model based on pre-lockdown out-of-sample performance, we would select the model with the 24 lagged hourly variables and 24 lead hourly variables. This model also yields a larger average predicted BAU re-dispatch costs during the lockdown. Therefore the actual average re-dispatch costs relative to the average predicted BAU re-dispatch costs during the lockdown are lower (+11\% compared to +37\%). The prediction errors of both dynamic model specifications imply Wilcoxon signed rank statistics indicating statistically significantly different distributions of predicted BAU versus actual re-dispatch costs. Hence, the qualitative interpretation of our results remain unchanged for these dynamic models to predict re-dispatch costs.

\begin{table}[ht] 
  \centering
  \caption{Model Comparison based on Pre-Lockdown Out-of-sample Performance}
    \begin{threeparttable}
    \begin{tabular}{cccrc}
    \toprule
    Polynomials/Interactions\tnote{1} & Lags\tnote{2} & Leads\tnote{3} & RMSE\tnote{4}& $\frac{\text{Actual cost}}{\text{Predicted BAU cost}}\tnote{5}$\\
    \midrule
    \multicolumn{5}{c}{\textit{Panel A}: Linear regression model} \\
         -- &  --     &     --  & 103,120  & +82\%\\
    2$^\text{nd}$ degree &   --    &    --   & 85,904 & +55\%\\
    2$^\text{nd}$ and 3$^\text{rd}$ degree &    --   & --      & 104,505 & +80\%\\ \addlinespace[5pt]
    \multicolumn{5}{c}{\textit{Panel B}: Deep neural network} \\
         -- & --      &   --    & 81,676 & +37\%\\
         -- & 24 hours     &  --     & 80,863 & +17\% \\
         -- & 24 hours     & 24 hours     & 80,363 & +11\%\\
    \bottomrule
    \end{tabular}%
    \begin{tablenotes}
    \item \tnote{1} $n^{\text{th}}$ degree polynomials and interaction terms of continuous explanatory variables, i.e., zonal net demands, zonal net demand forecasts, and day-ahead market price. 
    \item \tnote{2} Lags of system net demand.
    \item \tnote{3} Leads of system net demand forecast.
    \item \tnote{4} $\text{RMSE}(y,\hat{y}) = \sqrt{T^{-1} \sum_{t=1}^T (y_t - \hat{y}_t)^2}.$
    \item \tnote{5} Average actual hourly re-dispatch costs during lockdown relative to predicted BAU re-dispatch costs.
    \end{tablenotes}
    \end{threeparttable}
  \label{tab:perf}%
\end{table}%

\subsubsection{Renewables and the Cost to Re-dispatch}\label{sec:renew_and_redis}
One concern with our preferred results is that our predicted re-dispatch cost increase during the lockdown may be a lower bound on re-dispatch cost increase associated with the equivalent decrease in net demand from a larger share of intermittent renewables. That is because the lockdown demand shock may have been more evenly distributed (over time and space) than it would be the case if the share of intermittent renewables increased substantially. We are not able to test this hypothesis directly because the lockdown re-dispatch cost observations resulted from a negative demand shock. However, we can analyze how the effect of an increased share of solar and wind output would change the cost to re-dispatch the system using our BAU prediction model.\footnote{We refer to \cite{GiPa18} for a graphical analysis detailing how the increase in capacity from renewables may have increased the real-time re-dispatch cost in Italy's northern bidding zone comparing the years 2006--2008 (almost no renewable capacity) to the years 2013--2015. \cite{GiPa16} calculate price premia between the Italian day-ahead (intra-day) market prices and awarded quantity-weighted average real-time re-dispatch market payments and find that renewables generally increase those. \cite{BiBoRES16} analyzes zonal Lerner indices during the period 2009 to 2013 for the main generators in the Italian day-ahead market and conclude that the exercise of market power in the day-ahead market has been surprisingly reinforced in specific off-peak hours as a result from increased supply from renewables.} More precisely, we conduct three counterfactuals with different assumptions on how the increased output of renewables will be distributed over time and space. The overall reduction of zonal net demands is the same under all three scenarios.

We use the dynamic model presented in Section~\ref{sec:res_dynModel} with the only exception that we replace the day-ahead market price by the daily natural gas price to account changes in the input fuel costs. There exists an ample literature on how more output from low variable cost renewables will depress day-ahead market prices \citep[see e.g.,][]{SeRa08, WuLa13, ClCa15, SaCo20} and on how more output from renewables will affect the intra-day variance in hourly day-ahead market prices \citep[see e.g.,][]{WoGr16}. Hence, when manipulating the output from renewables we cannot treat the day-ahead market price as exogenous variable any longer and therefore replace it by the daily gas price assuming that this price is unaffected by a modified aggregate renewables output profile. We believe that applying a dynamic model is important in this particular setting because especially solar with its unimodal output distribution will likely change the operation of many conventional units.

The first counter-factual analysis involves scaling the existing locational (zonal) output from wind and solar by factor two.\footnote{Table~\ref{tab:resCap}, Panel B, shows several projections on the growth of wind and solar capacity in Italy and find that wind and solar capacity may already be doubled in the next decade from now.} The simplifying assumptions behind this scaling approach is that the frequency of binding intra-zonal transmission constraints will not increase. Furthermore, we assume that weather patterns within a zone are comparable and that the zonal distribution of additional renewable generation capacity will be equal to the current distribution. All other factors that define the zonal net demands and day-ahead forecasts are unchanged. 

We use our deep neural network model trained and validated on 2017 to 2019 data to predict the hourly re-dispatch costs between 2017-01-01 and 2020-03-07 for the net demand implied by our counterfactual renewable output. Table~\ref{tab:cf} shows that doubling the output from solar and wind generation units leads to an increase in the predicted re-dispatch cost by 37\% (Table~\ref{tab:cf}, first row). This percent increase is larger than the percent increase that predicted lockdown re-dispatch costs are relative to our counterfactual BAU lockdown period re-dispatch costs from past years (26\%). One reason for this deviation may be that the larger sample length is more informative and portrays a better overall picture. Another potential explanation is that the variable output from renewables has greater impact on re-dispatch costs than a persistent demand shock---a hypothesis we investigate in the next paragraphs.

The second counter-factual analysis aims to quantify the effect of intermittency and temporal variation in the output of renewables. We therefore reduce the zonal net demands and net demand forecasts uniformly by the sample average output of wind and solar units in each zone.\footnote{We calculate the average actual output of solar and wind for each zone and net this value from each zone's net demand.} This exercise yields a predicted increase in the re-dispatch costs of 32\% (Table~\ref{tab:cf}, second row) during period 2017-01-01 and 2020-03-07.

Lastly, we modify the spatial distribution smoothing the additional renewables along time and space.\footnote{We distributed the additional output from renewables uniformly across the six demand zones. Put differently, we calculate the average aggregate output of solar and wind units and net one sixth of this value from the zonal net demands.} This exercise yields a predicted increase in the hourly re-dispatch cost by only 25\% (Table~\ref{tab:cf}, third row) for the period 2017-01-01 to 2020-03-07.

These findings emphasize the importance of the various seasonalities of renewable technologies, the importance of their location, as well as the associated forecast errors between day-ahead prediction and real-time output of renewables.  These final two estimates emphasize that increased spatial and temporal variation in renewables output as the share of renewables output in the system increases is likely to result in additional re-dispatch cost increases that cannot be explained by historic observations. Therefore these estimates of the impact of the doubling renewables output are likely to be conservative.

\begin{table}[ht] 
  \centering
  \caption{Predicted BAU Re-Dispatch Cost doubling Output from Renewables}
  \begin{threeparttable}
    \begin{tabular}{cccrr}
    \toprule
    Solar output & Wind output & Smoothed & $\Delta \overline{\text{Cost}}$ [EUR/hour] & Relative Change \\
    \midrule
    double & double &  --    & +69,845  & +37\% \\
    double & double &  time     & +60,140  & +32\% \\
    double & double &  time/space\tnote{1}     & +47,380  & +25\% \\
    \bottomrule
    \end{tabular}%
    \begin{tablenotes}
    \item \tnote{1} Distributed uniformly across the six demand zones.
    \item \textit{Notes}: Modeled trained and validated on 2017 to 2019 data and evaluated during the period January 1, 2017 to March 7, 2020. Average hourly re-dispatch cost during this period was 189,000\;EUR. Deep neural network model including leads and lags as described in Section~\ref{sec:res_dynModel} deployed to produce predictions.
    \end{tablenotes}
    \end{threeparttable}
  \label{tab:cf}%
\end{table}%

\section{Discussion and Conclusion}\label{sec:conclusion}
We use the negative demand shock to the Italian electricity market as a result of the COVID-19 lockdown to study the impact of an increase in renewable generation capacity in the Italian electricity market on re-dispatch costs. We find that the COVID-19 demand shock yields same reduction in the demand for energy from controllable generation units than a slightly more than doubling of renewable energy production would under business-as-usual demand conditions.

Using a neural network model to provide a realistic business as usual predictive model for re-dispatch costs, we compute counterfactual re-dispatch costs for the first four months in 2020 in Italy and compare these predicted re-dispatch costs to actual re-dispatch costs for the pre-COVID-19 lockdown and COVID-19 lockdown periods.

We find no statistical difference between the distribution of predicted hourly re-dispatch costs and actual re-dispatch costs during the pre-COVID-19 period. For the COVID-19 period, actual re-dispatch costs are 37\% higher than predicted business-as-usual re-dispatch costs. Blowing up this increase in re-dispatch costs during the COVID-19 period to an annual value and using the fact that the demand reduction implies a doubling of renewable energy production implies a roughly 1 billion EUR annual increase in re-dispatch costs associated with doubling renewable output. We emphasize that this re-dispatch cost increase is not a prediction of how these costs will scale with an increased wind and solar generation share. They are only indicative of how much these costs could increase without system operators making the necessary investments in transmission and other technologies to manage the lower levels of net demand that result from an increased share of intermittent renewables.

There are several reasons why a substantial increase in the share of renewables may have an even more severe effect on re-dispatch costs in the Italian market than a negative demand shock. First, there is a difference in expected shape and location to net demand from a reduction in gross demand versus an increase in intermittent renewables. In the case of solar, an energy-equivalent increase would have a far more concentrated diurnal (and seasonal) shape than an energy-equivalent decrease in gross demand, occurring across all hours. For wind, the diurnal shape may not be as concentrated, but the location is. In both of these situations---the more concentrated shape and concentrated location---one could reasonably expect higher re-dispatch costs from the increase in renewables, as compared to the decrease in gross demand. Second, going beyond hourly averages to hourly variances, there is likely to be a big difference in the volatility of hourly net demand from an increase in renewables versus a decrease in gross demand due to the former's intermittency. We have shown that both effects are present in our analyses of different counterfactual renewable output increases in Section~\ref{sec:robustness}. However, as we learned from our lockdown period analysis in Section~\ref{sec:res} an \textit{additional} re-dispatch cost increase is likely as result of persistently lower net demands, so our estimate of the increase of re-dispatch costs is on the conservative side.

We should also note that our analysis points out the need for market power mitigation mechanisms to deal with a new source of local market power---that due low levels of net demand. Traditional market power mitigation mechanisms focus on high demand hours as these were the hours were little supply capacity will be left to compete with each other to serve demand. According to this logic, low demand hours are not typically thought to be periods when suppliers can exercise unilateral market power because there is plenty of idle generation capacity. However, commitment costs, system security constraints, or transmission constraints are the reasons for market power potential in low demand hours. Although grid upgrades can help in relieving these system security constraints and, consequently, the opportunities to exercise local market power, these investments can have long lead-times (ranging from few years, in the case of devices located inside substations, to decades, in the case of transmission lines) and in a dynamic environment it is hard to anticipate what will be needed in years to decades from now. Therefore, dynamic on-line market power mitigation systems could be useful to mitigate high re-dispatch costs as seen during the lockdown. Such systems have the capacity to properly mitigate local market power even when the power system is affected by unexpected events.

\singlespacing
\bibliography{lit}

\clearpage
\appendix
\counterwithin{figure}{section}
\counterwithin{table}{section}
\doublespacing

\section{Data Description} \label{app:data}
We use hourly data that spans between 2017-01-01 and 2020-04-26. In Table~\ref{tab:data}, we detail input data as well as their sources. The dependent variable ($y$) is the total hourly real-time re-dispatch cost in the Italian electricity market.\footnote{Re-dispatch costs only include the costs to change the schedules from dispatchable units but do not include the costs to start up or to change a unit's configuration.} All other market data is used to compute zonal net demands and net demand forecasts for dispatchable supply as described in Equation~\ref{eq:rd}. Our final regressor matrix contains zonal net demands, zonal net demand forecasts, an indicator variable for workdays, and an indicator variable for winter. The latter is important, as in winter more electricity can be transported through the existing transmission network because outside temperatures are lower.

In Table~\ref{tab:desc}, we display the mean, standard deviation (Std), minimum (Min), and maximum (Max) of each of the variables using our predictive modeling exercise.

\begin{table}[ht] 
  \centering
  \footnotesize
  \caption{Data Description}
  \begin{threeparttable}
    \begin{tabular}{p{4cm}lp{2cm}p{2cm}p{4cm}}
    \toprule
    Variable & Unit  & Temporal resolution & Spatial resolution & Comment\\
    \midrule
    Re-dispatch cost\tnote{1} & EUR   & Hourly & National & Excluding start-up costs and configuration change costs \\ \addlinespace[5pt]
    Day-ahead Market Price\tnote{2}  & EUR/MWh & Hourly & National & Relevant for demand side (PUN)\tnote{3}\\
    Gas price\tnote{2}	& EUR/MWh	& Daily	& National & Average day-ahead market price at the Italian virtual trading hub, PSV (Punto di Scambio Virtuale)\tnote{4} \\
    Market demand\tnote{5} & MWh   & Hourly & Zonal\tnote{7} & Actual total demand minus self-consumption that includes industrial self-consumption and distributed solar \\
    Market demand forecast\tnote{5} & MWh   & Hourly & Zonal\tnote{7} & Day-ahead forecast total demand minus self-consumption that includes industrial self-consumption and distributed solar \\
    Solar generation\tnote{6} & MWh   & Hourly & Zonal\tnote{7} & Actual solar generation \\
    Solar generation forecast\tnote{6} & MWh   & Hourly & Zonal\tnote{7} & Day-ahead forecast solar generation \\
    Wind generation\tnote{6} & MWh   & Hourly & Zonal\tnote{7} & Actual wind generation \\
    Wind generation forecast\tnote{6} & MWh   & Hourly & Zonal\tnote{7} & Day-ahead forecast wind generation \\
    Hydro generation\tnote{6} & MWh   & Hourly & Zonal\tnote{7} & Actual non-controllable hydro generation (run-of-river)\\
    Net imports\tnote{5}  & MWh   & Hourly  & Neighboring countries\tnote{8} & Net imports to Italy from neighbouring countries \\ \addlinespace[5pt]
    Workday &     &     &     & Indicator variable equal to one for days that are not weekend days nor holidays \\
    Winter &    &     &     & Indicator variable equal to one for months October to April \\
    \bottomrule
    \end{tabular}%
    \begin{tablenotes}
    \item \tnote{1} Data provided by Terna.
    \item \tnote{2} Data publicly available from \url{https://www.mercatoelettrico.org}.
    \item \tnote{3} We use the prezzo unico nazionale (PUN) that is effectively the demand-weighted average zonal price relevant for the demand side \citep[see][for more details]{GrWo20}.
    \item \tnote{4} Continuous trading market.
    \item \tnote{5} Data publicly available from \url{https://www.terna.it/en/electric-system/transparency-report}.
    \item \tnote{6} Data publicly available from \url{https://transparency.entsoe.eu}.
    \item \tnote{7} Italian bidding zones considered are North, Center-North, Center-South, South, Rossano, Sardinia, and Sicily. In 2019, Brindisi and Foggia bidding zones were integrated into South and Priolo bidding zone was integrated into Sicily.
    \item \tnote{8} Italy is connected to Austria (North), France (North; Sardina and Center-North is connected via Corsica), Greece (South), Malta (Sicily), Montenegro (Center-South), Slovenia (North), and Switzerland (North).
    \end{tablenotes}
    \end{threeparttable}
  \label{tab:data}%
\end{table}%

\begin{table}[ht] 
  \centering
  \footnotesize
  \caption{Descriptive Statistics}
    \begin{threeparttable}
    \begin{tabular}{lllrrrrr}
    \toprule
    Variable & Unit & Location & \multicolumn{1}{c}{Count} & \multicolumn{1}{c}{Mean} & \multicolumn{1}{c}{Std} & \multicolumn{1}{c}{Min} & \multicolumn{1}{c}{Max} \\
    \midrule
    Re-dispatch cost & EUR & National & 29,087 & 197,362 & 132,602 & $-107,147$ & 1,328,790 \\ \addlinespace[5pt]
    Day-ahead market price & EUR/MWh & National & 29,087 & 54    & 16    & 0     & 170 \\ 
    Gas Price &	EUR/MWh	& National &	29,087 &	19	& 5	& 8	& 56 \\ \addlinespace[5pt]
    Net demand & MWh & North & 29,087 & 9,930 & 3,964 & 561   & 23,633 \\
    Net demand forecast & MWh & North & 29,087 & 10,189 & 3,973 & 1,129 & 24,036 \\ \addlinespace[5pt]
    Net demand & MWh& Center-North & 29,087 & 2,970 & 749   & 545   & 5,433 \\
    Net demand forecast & MWh & Center-North & 29,087 & 3,044 & 754   & 652   & 5,483 \\ \addlinespace[5pt]
    Net demand & MWh & Center-South & 29,087 & 3,892 & 1,112 & 987   & 7,714 \\
    Net demand forecast & MWh & Center-South & 29,087 & 3,992 & 1,116 & 655   & 7,832 \\ \addlinespace[5pt]
    Net demand & MWh & South & 29,087 & 1,523 & 1,110 & $-1,979$ & 4,672 \\
    Net demand forecast & MWh & South & 29,087 & 1,668 & 1,082 & $-2,534$ & 4,691 \\ \addlinespace[5pt]
    Net demand & MWh & Sardinia\tnote{1} & 29,087 & 722   & 299   & $-559$  & 1,532 \\
    Net demand forecast & MWh & Sardinia\tnote{1} & 29,087 & 734   & 297   & $-606$  & 1,598 \\ \addlinespace[5pt]
    Net demand & MWh & Sicily & 29,087 & 1,520 & 534   & $-163$  & 3,550 \\
    Net demand forecast & MWh & Sicily & 29,087 & 1,565 & 528   & $-580$  & 3,718 \\ \addlinespace[5pt]
    Net demand & MWh & Rossano\tnote{2} & 29,087 & $-39$   & 39    & $-180$  & 0 \\
    Net demand forecast & MWh & Rossano\tnote{2} & 29,087 & $-40$   & 36    & $-167$  & 0 \\
    \bottomrule
    \end{tabular}%
    \begin{tablenotes}
    \item \tnote{1} Net imports from Corsica are only considered in Sardinia's net demand and net demand forecast.
    \item \tnote{2} Limited production bidding zone with no demand.
    \item \textit{Notes}: The data spans the period from January 1, 2017 to April 26, 2020. It is hourly data except for the gas price which is available on a daily basis. Missing values in the gas price time series and in any of the time series used to construct zonal net demands and zonal net demand forecasts are linearly interpolated.
    \end{tablenotes}
    \end{threeparttable}
  \label{tab:desc}%
\end{table}%

\section{Research Design}\label{app:resDes}
\begin{figure}[ht]
\caption{Research Design}
    \includegraphics[width=1 \textwidth]{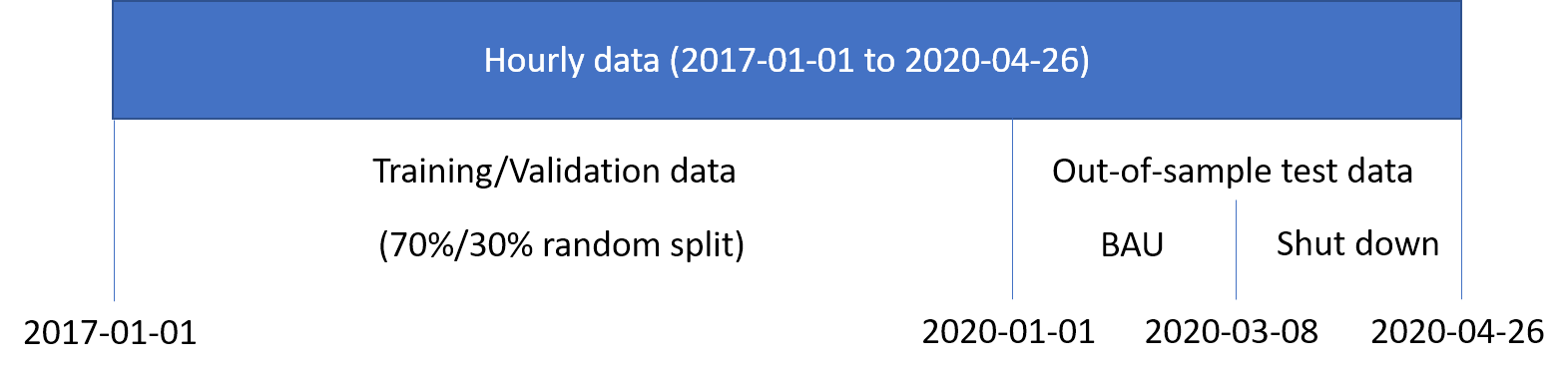}
    \label{fig:resDesign}
\end{figure}

In Figure~\ref{fig:resDesign}, we summarize the research design. After collecting the hourly data on the variables described in Table~\ref{tab:data}, we separate the data in a training and validation dataset and in an out-of-sample dataset. The out-of-sample dataset contains data from the beginning of the year 2020 until April 26, 2020 which was the day when the lockdown was eased. We divide the out-of-sample dataset in a pre-lockdown period that lasts from January 1, 2020 until March 7, 2020 and in a lockdown period that covers the rest of the out-of sample data. The in-sample data spans from January 1, 2017 to December 31, 2019 and we randomly take 70\% of the days in the sample as training data and the remaining 30\% as validation data. We train the deep learning model on the training data and use the validation data to avoid over-fitting of the model. Out-of-sample data is completely set-aside data that never has been presented to the algorithm.

We use the trained weights to predict hourly out-of-sample re-dispatch costs. The prediction error of our model is calculated using out-of-sample pre-lockdown data. In a last step, we calculate predicted BAU re-dispatch costs including a 95\% prediction band. These predicted BAU re-dispatch costs are then compared to the actual re-dispatch cost realizations during the lockdown and observations that lie outside of the prediction band are considered to be excessive and associated with the special low-demand period. 

\section{Deep Neural Network Configuration and Performance} \label{app:dl}
We train a plain vanilla \textit{deep neural network} with two hidden layers. We minimize the mean squared error (MSE) given by $T^{-1}\sum_{t=1}^T ( y_t - \hat{y}_t )^2$, where $y~=~(y_1 , y_2 , ..., y_T )'$ is the vector of \textit{actual total re-dispatch costs} and $\hat{y}~=~(\hat{y}_1 , \hat{y}_2 ,..., \hat{y}_T )'$ the vector of their predictions. We use Root Mean Square Propagation (RMSProp) as an optimizer that is a stochastic gradient descent algorithm in which the learning rate is adapted for each of the parameters.

We set the maximum value of epochs to 1,500, but apply ``early stopping'' of the optimization routine to avoid over-fitting \citep{YaRo07}. More precisely, we stop the routine if the accuracy in the validation set worsens for five consecutive epochs. Optimizing hyper-parameters  applying a learning rate equal to 0.006, 48 neurons ($n$) in each of the two hidden layers (makes 3,265 trainable parameters for our preferred model specification), and use of a rectified linear unit (ReLU) activation function. We standardize each explanatory variable by removing the mean and scaling to unit variance. Mean and variance of the training data were applied to standardize the validation data. We account for model uncertainty by using dropout that randomly switches off a fraction of the neurons in the neural network. This technique aims at reducing over-fitting and to improve training performance. The hyper-parameter optimization yields the optimal dropout fraction to be 0.1. The model is trained using the libraries \textit{Tensorflow~2.1.0} and \textit{Keras~1.0.1}. We use \textit{hyperband} as a tuner to perform hyper-parameter optimization \citep{LiJa18}. Optimal hyper-parameters and the pre-defined set of hyper-parameters are presented in Table~\ref{tab:dlPara}. Note that hyper-parameters are optimized using only training and validation data but not out-of-sample data.

The RMSE of our preferred specification of the deep neural network using out-of-sample pre-lockdown observations and predictions yields 81,676\;EUR.

\begin{table}[ht]
  \centering
  \caption{Optimal Hyper-parameters}
    \begin{tabular}{lrr}
    \toprule
    Hyper-parameter & \multicolumn{1}{c}{Optimal value} & \multicolumn{1}{c}{Possible values}\\
    \midrule
    Activation function & ReLU & ReLU, Tanh, Sigmoid\\
    Learning rate & 0.006161 & Min: 1e-4; Max: 1e-2; ``log-sampling''\\
    Dropout  & 0.1 & 0.1, 0.2, \ldots, 0.5\\
    $n_1$ & 48 & 16, 32, 48\\
    $n_2$ & 48 & 16, 32, 48\\
    \bottomrule
    \end{tabular}%
  \label{tab:dlPara}%
\end{table}%

\section{Additional Figures} \label{app:addFig}
\begin{figure}
\caption{Hourly average Re-dispatch Cost}
    \centering
	\includegraphics[width=.5 \textwidth]{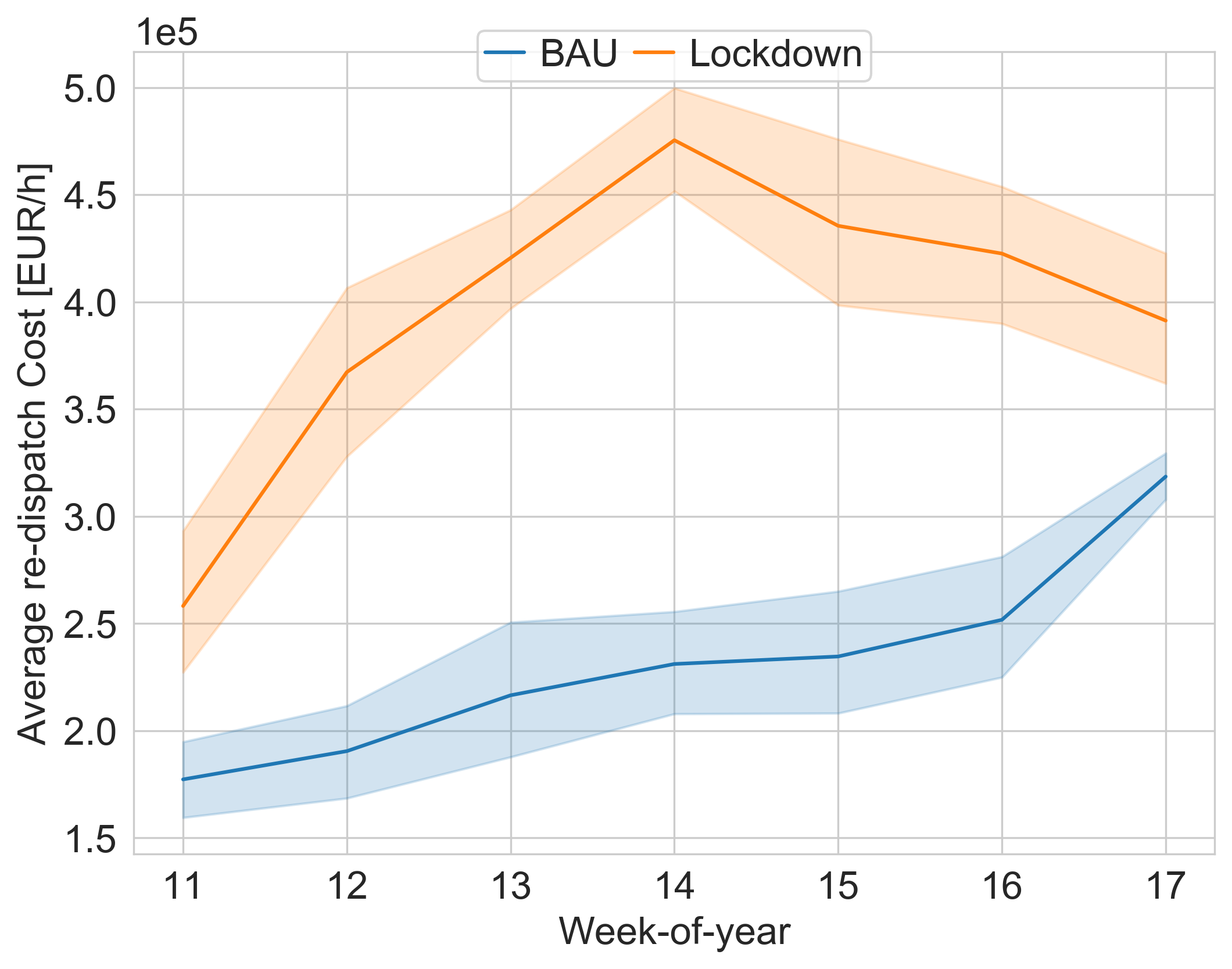}
	\floatfoot{\footnotesize \textit{Notes:} Business-as-usual (BAU) re-dispatch costs calculated as the average cost for each hour of the week over March and April during the years 2017--2019. Lockdown re-dispatch costs calculated as the average hourly cost between March 9, 2020 and April 26, 2020.  Shaded area around the mean values represents the 95\% confidence interval. Data sources to derive the graph described in Table~\ref{tab:data}.}
    \label{fig:reDispatchCost}
\end{figure}

Figure~\ref{fig:reDispatchCost} compares weekly BAU re-dispatch costs averaged over all hours in a week to lockdown re-dispatch cost. BAU re-dispatch costs calculated as the average cost for each hour of the week over March and April during the years 2017--2019. Lockdown re-dispatch costs are calculated as the average hourly cost between 2020-03-09 and 2020-04-26.

\begin{figure}[ht]
\caption{Net demand shock due to Lockdown as a Response to COVID-19 Pandemic}
    \centering
	\includegraphics[width=.8 \textwidth]{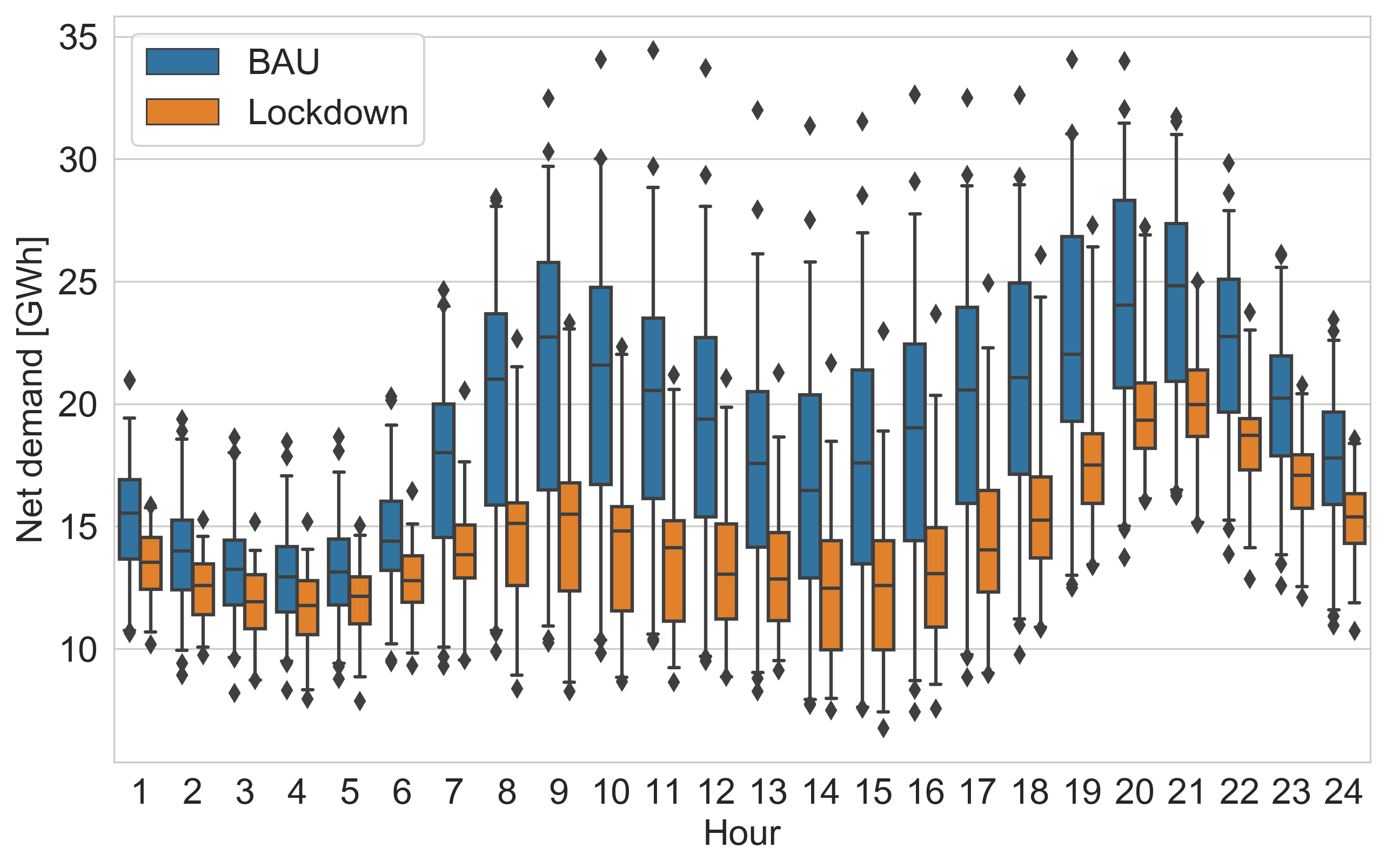}
    \floatfoot{\footnotesize \textit{Notes:} Business-as-usual (BAU) system hourly net demand defined as the sum of zonal net demands defined in Equation~\ref{eq:rd}. BAU system net demands for each hour in March and April for the years 2017 to 2019. Lockdown net demands for each hour between March 8, 2020 and April 26, 2020. Boxes represent interquartile range (IQR) and upper and lower vertical bars equal to the 1 percent and 99 percent. Diamonds represent outliers not included in the 1--99 percentile. Data sources to derive the graph described in Table~\ref{tab:data}.}
    \label{fig:shockRd}
\end{figure}

Figure~\ref{fig:shockRd} compares hourly system BAU net demands to system net demands under lockdown.\footnote{System net demand is defined as the sum over zonal net demands. Zonal net demands are defined in Equation~\ref{eq:rd}.} BAU net demands for each hour in March and April for the years 2017 to 2019. Lockdown net demands for each hour between 2020-03-08 and 2020-04-26. Boxes represent interquartile range (IQR) and upper and lower vertical bars equal to the 1 percent and 99 percent. Diamonds represent outliers not included in the 1--99 percentile.

\end{document}